\documentclass{pasj02}
\Received{$\langle$reception date$\rangle$}
\Accepted{$\langle$acception date$\rangle$}
\Published{$\langle$publication date$\rangle$}
\usepackage[switch,mathlines]{lineno}

\begin{document}

\title{Soft X-ray Imager of the Xtend system onboard \textit{XRISM}
\thanks{The corresponding authors are Hirofumi Noda, Koji Mori, Hiroshi Tomida, Hiroshi Nakajima, Takaaki Tanaka, and Hiroshi Murakami.
}
}
\author{Hirofumi Noda$^{1}$, Koji Mori$^{2,3}$, Hiroshi Tomida$^{3}$, Hiroshi Nakajima$^{4,3}$, Takaaki Tanaka$^{5}$, Hiroshi Murakami$^{6}$, Hiroyuki Uchida$^{7}$, Hiromasa Suzuki$^{3}$, Shogo Benjamin Kobayashi$^{8}$, Tomokage Yoneyama$^{9}$, Kouichi Hagino$^{10}$, Kumiko Nobukawa$^{11}$, Hideki Uchiyama$^{12}$, Masayoshi Nobukawa$^{13}$, Hironori Matsumoto$^{14}$, Takeshi Go Tsuru$^{7}$, Makoto Yamauchi$^{2}$, Isamu Hatsukade$^{2}$, Hirokazu Odaka$^{14}$, Takayoshi Kohmura$^{15}$, Kazutaka Yamaoka$^{16}$, Tessei Yoshida$^{3}$, Yoshiaki Kanemaru$^{3}$, Junko Hiraga$^{17}$, Tadayasu Dotani$^{3}$, Masanobu Ozaki$^{18}$, Hiroshi Tsunemi$^{14}$, Jin Sato$^{2}$, Toshiyuki Takaki$^{2}$, Yuta Terada$^{2}$, Keitaro Miyazaki$^{2}$, Kohei Kusunoki$^{2}$, Yoshinori Otsuka$^{2}$, Haruhiko Yokosu$^{2}$, Wakana Yonemaru$^{2}$, Kazuhiro Ichikawa$^{2}$, Hanako Nakano$^{2}$, Reo Takemoto$^{2}$, Tsukasa Matsushima$^{2}$, Reika Urase$^{2}$, Jun Kurashima$^{2}$, Kotomi Fuchi$^{2}$, Kaito Hayakawa$^{4}$, Masahiro Fukuda$^{4}$, Takamitsu Kamei$^{4}$, Yoh Asahina$^{4}$, Shun Inoue$^{7}$, Amano Yuki$^{3}$, Yuma Aoki$^{11}$, Yamato Ito$^{11}$,  Tomoya Kamatani$^{11}$, Kouta Takayama$^{11}$, Takashi Sako$^{13}$, Marina Yoshimoto$^{14}$, Kohei Shima$^{14}$, Mayu Higuchi$^{15}$, Kaito Ninoyu$^{15}$, Daiki Aoki$^{15}$, Shun Tsunomachi$^{15}$, and Kiyoshi Hayashida$^{14}$}%
\altaffiltext{1}{Astronomical Institute, Tohoku University, 6-3 Aramakiazaaoba, Aoba-ku, Sendai, Miyagi 980-8578, Japan}
\altaffiltext{2}{Faculty of Engineering, University of Miyazaki, 1-1 Gakuen Kibanadai Nishi, Miyazaki, Miyazaki 889-2192, Japan}
\altaffiltext{3}{Japan Aerospace Exploration Agency, Institute of Space and Astronautical Science, 3-1-1 Yoshino-dai, Chuo-ku, Sagamihara, Kanagawa 252-5210, Japan}
\altaffiltext{4}{College of Science and Engineering, Kanto Gakuin University, Kanazawa-ku, Yokohama, Kanagawa 236-8501, Japan}
\altaffiltext{5}{Department of Physics, Konan University, 8-9-1 Okamoto, Higashinada, Kobe, Hyogo 658-8501}
\altaffiltext{6}{Faculty of Informatics, Tohoku Gakuin University, 3-1 Shimizukoji, Wakabayashi-ku, Sendai, Miyagi 984-8588}
\altaffiltext{7}{Department of Physics, Kyoto University, Kitashirakawa Oiwake-cho,Sakyo-ku, Kyoto, Kyoto 606-8502, Japan}
\altaffiltext{8}{Department of Physics, Faculty of Science, Tokyo University of Science, Kagurazaka, Shinjuku-ku, Tokyo 162-0815, Japan}
\altaffiltext{9}{Faculty of Science and Engineering, Chuo University, 1-13-27 Kasuga, Bunkyo-ku, Tokyo 112-8551, Japan}
\altaffiltext{10}{Department of Physics, University of Tokyo, 7-3-1 Hongo, Bunkyo-ku, Tokyo 113-0033, Japan}
\altaffiltext{11}{Department of Physics, Kindai University, 3-4-1 Kowakae, Higashi-Osaka, Osaka 577-8502, Japan}
\altaffiltext{12}{Science Education, Faculty of Education, Shizuoka University, Suruga-ku, Shizuoka, Shizuoka 422-8529, Japan}
\altaffiltext{13}{Faculty of Education, Nara University of Education, Nara, Nara 630-8528, Japan}
\altaffiltext{14}{Department of Earth and Space Science, Osaka University, 1-1 Machikaneyama-cho, Toyonaka, Osaka 560-0043, Japan}
\altaffiltext{15}{Department of Physics, Faculty of Science and Technology, Tokyo University of Science, 2641 Yamazaki, Noda, Chiba 270-8510, Japan}
\altaffiltext{16}{Department of Physics, Nagoya University, Chikusa-ku, Nagoya, Aichi 464-8602, Japan}
\altaffiltext{17}{Department of Physics, Kwansei Gakuin University, 2-2 Gakuen, Sanda, Hyogo 669-1337, Japan}
\altaffiltext{18}{Advanced Technology Center, National Astronomical Observatory of Japan, Mitaka, Tokyo 181-8588, Japan}
\email{hirofumi.noda@astr.tohoku.ac.jp}


\maketitle

\begin{abstract}
The Soft X-ray Imager (SXI) is  the X-ray charge-coupled device (CCD) camera for the soft X-ray imaging telescope Xtend installed on the \textit{X-ray Imaging and Spectroscopy Mission} (\textit{XRISM}), which was  adopted as a recovery mission for the \textit{Hitomi} X-ray satellite and was successfully launched on 2023 September 7 (JST).
In order to maximize the science output of  \textit{XRISM}, we set the requirements for Xtend and find that the CCD set employed in the \textit{Hitomi}/SXI or similar, i.e., a $2 \times 2$ array of back-illuminated CCDs with a $200~\mu$m-thick depletion layer, would be practically best among available choices, when used in combination  with the X-ray mirror assembly.
We design the \textit{XRISM}/SXI, based on the \textit{Hitomi}/SXI, to  have a wide field of view of $38' \times 38'$ in the 0.4--13~keV energy range.  
  We incorporated several  significant improvements from the \textit{Hitomi}/SXI into the CCD chip design to enhance the optical-light blocking capability and to increase the cosmic-ray tolerance, reducing the degradation of  charge-transfer efficiency  in orbit. 
 By the time of the launch of \textit{XRISM}, the imaging and spectroscopic capabilities of the SXI  has been extensively studied  in on-ground experiments  with the full flight-model configuration or equivalent setups and confirmed to meet the requirements. 
The optical blocking capability, the cooling and temperature control performance,  and the transmissivity and quantum efficiency to incident X-rays of the CCDs are also all confirmed to meet the  requirements.
 Thus, we successfully complete the pre-flight development of the SXI for \textit{XRISM}.
\end{abstract}
%\pagewiselinenumbers

%============S1============
\section{Introduction}

%============S1============

\textit{X-ray Imaging and Spectroscopy Mission} (\textit{XRISM}) (\cite{Tashiro2018}; \cite{Tashiro2020}; \cite{Tashiro2024}) is the seventh Japanese X-ray astronomical satellite,  adopted as the recovery mission of the short-lived \textit{ASTRO-H} (\textit{Hitomi}) satellite \citep{Takahashi2018}.
 Its development commenced in  2017, and it was successfully launched by the H-IIA rocket F47 from the JAXA Tanegashima Space Center on 2023 September 7 (JST).
Although its predecessor \textit{Hitomi} was lost about a month after its launch on 2016 February 17 due to attitude-control troubles, it  demonstrated during its short lifetime the power of ultra-precise X-ray spectroscopy on multiple astrophysical objects  with  an X-ray microcalorimeter for the first time. 
 Most notably, the observations of the Perseus cluster of galaxies revealed the detailed velocity and temperature distributions, heavy element composition of its hot plasma, and structure of an active galactic nucleus (AGN) at its center with an unprecedented precision 
 (e.g., \cite{HitomiFirst}; \cite{HitomiAbundance}; \cite{HitomiVelocity}; \cite{HitomiTemperature}; \cite{HitomiAGN}). 
Considering the  significance and uniqueness of  X-ray spectroscopy of cosmic sources in an ultra-high energy resolution, it was first decided that \textit{Hitomi}'s recovery mission, \textit{XRISM}, should be equipped with a soft X-ray spectroscopy telescope ``Resolve,'' a system with an X-ray microcalorimeter combined with thin foil-nested conically approximated Wolter-I mirror optics with a focal length of 5.6~m (X-ray Mirror Assembly; XMA).  

 The downside of Resolve  is a relatively narrow field of view (FOV) of $3'\times 3'$ and a large pixel size of $30''$, which is larger than the size at the core of the Point-Spread Function (PSF) of the XMA.  The mission team  decided to employ on \textit{XRISM} a complementary  instrument for wider-FOV, finer, and more precise imaging in the same energy band.
 For this purpose, a wide  FOV  X-ray imager with a small pixel size and low Non X-ray Background (NXB) level is essential.  Accordingly, it was decided that the soft X-ray imaging telescope ``Xtend,'' a system with  an X-ray charge-coupled device (CCD) camera named ``Soft X-ray Imager'' (SXI) combined with  a separate XMA from that for Resolve  would  also be installed on \textit{XRISM} (namely, Xtend = SXI + XMA; \cite{Hayashida2018}; \cite{Nakajima2020}; \cite{Mori2022};  \cite{Mori2024}). The design of the SXI is based on the CCD camera of the predecessor \textit{Hitomi} (which was also called the SXI; see Section~2), but some significant improvements are made, as explained in detail in this paper. 

This paper  presents the description of the SXI of the Xtend system onboard \textit{XRISM} and its basic characteristics, 
 organized as follows. 
Section~2  gives the requirements for Xtend. 
Section~3  explains the design of the SXI  with a particular emphasis on design improvements from the \textit{Hitomi}/SXI.
Section~4 describes the on-board and on-ground event processing. 
Section~5 shows  the SXI's performances measured  with experiments with the full flight-model configuration and equivalent setups. 
Note that  the known issues of anomalous charge intrusion  on the  CCD chips of the same model but non-flight-model ones and the countermeasures for  the issues  are reported in the separate paper by \citet{Noda2024}. 
The in-orbit performance of the \textit{XRISM}/Xtend/SXI will be also reported in a separate paper.

%============S2============
\section{Requirements for Xtend}
%============S2============

%\begin{verbatim}
\begin{figure*}
 \begin{center}
  \includegraphics[width=140mm]{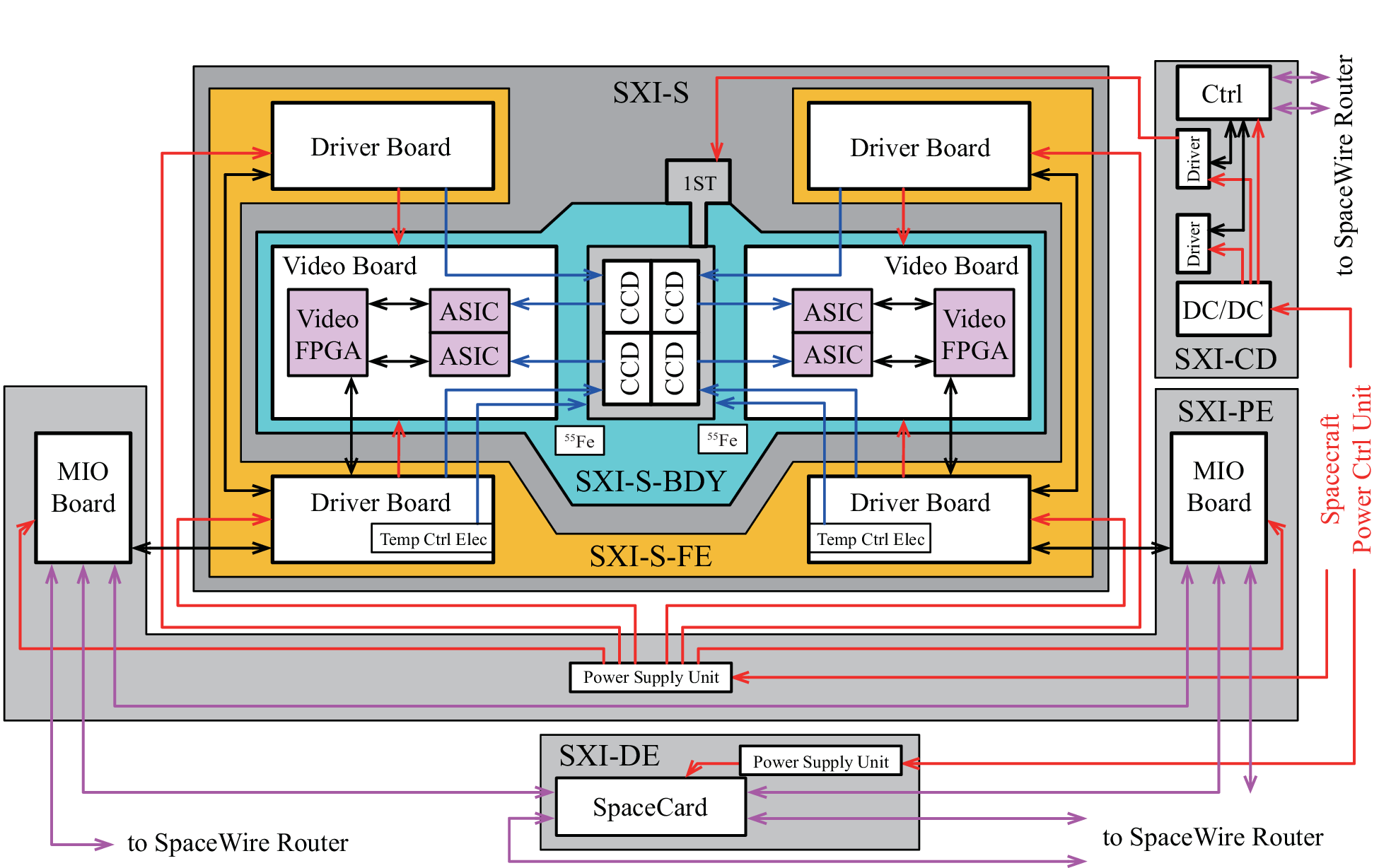}
 \end{center}
 \caption{Block diagram of the \textit{XRISM}/SXI system, modified from the diagram of the \textit{Hitomi}/SXI system in \citet{Tanaka2018}.  Red, blue, purple, and black lines show the power supply, analog, SpaceWire, and  notable digital signal flows, respectively. {Alt text: One block diagram.}}
 \label{fig1}
\end{figure*}
%\end{verbatim}

The following list summarizes the desired features of Xtend with the aim of  maximizing the science outputs by \textit{XRISM}. 
\begin{enumerate}
\item Enhancement of the potentials of spectroscopy in an unprecedentedly high~energy resolution   with Resolve, including the reduction of non-target-source contamination and noise
\begin{enumerate}
\item When a point source is present in the FOV of Resolve together with a diffuse source extending beyond the FOV of Resolve, (e.g., an AGN at the center of a cluster of galaxies), Xtend can distinguish their spectra by utilizing signals outside the FOV of Resolve. 
\item %When a bright transient appears outside the FOV of Resolve  part of the flux of  which  contaminates  the target signals, 
Even if a bright transient appears outside the FOV of Resolve and a part of its flux contaminates the main target signals within the FOV,
Xtend can evaluate and exclude the flux contamination precisely, taking account of its energy dependence. 
\item The combination of Xtend and Resolve enhances continuum-emission statistics   with an increase of a signal-to-noise ratio  by a factor of $\sim \sqrt{2}$ for a point source and up to an order of magnitude for a diffuse source   (while the gate valve of Resolve remains open\footnote{In the first year of operation of \textit{XRISM} in orbit (2023--2024), the gate valve of Resolve remained closed,which had maintained vacuum in the dewar on ground and was planned to be opened before the normal operation phase began. The gate valve in the closed position blocks a significant part of incoming X-rays to the X-ray microcalorimeter of Resolve, by $\sim 40$\% in effective area even at relatively-high 6~keV and much worse in the lower energy bands, up to virtually 100\% below 1.7~keV. As a result, Xtend/SXI's role has turned out to be even more significant than the pre-launch expectation, given that Resolve has so far had a much reduced effective area in the lower energy band.
}). The fact that the two detectors cover almost the same energy band is an advantage for this. 
\item In modeling highly extended diffuse sources beyond the field of view, be it in the foreground (e.g., Solar-wind charge exchange emission) or background (e.g., Galactic diffuse sources), Xtend's much wider FOV yields much higher statistics than those with Resolve, hence providing much higher statistics for their data.
\item Some operations and calibrations in orbit cannot be  completed by Resolve alone and require Xtend, such as the confirmation of a pointing direction, the calibration of the PSF, and the molecular contamination monitoring necessary prior to  the operation of opening the gate valve of Resolve. 
\end{enumerate}
\item Soft X-ray spectroscopic imaging with the wide FOV of Xtend with a low NXB level
\begin{enumerate}
%Xtend can effectively beused for studies of widely extended diffuse sources, such as the outer parts of clusters of galaxies, supernova remnants, Galactic ridge X-ray emission, superbubbles, and Warm-Hot Intergalactic Medium (WHIM). 
\item Xtend is particularly effective for studying widely extended diffuse sources, such as the outer regions of galaxy clusters, supernova remnants, Galactic ridge X-ray emission, superbubbles, and the Warm-Hot Intergalactic Medium (WHIM), outperforming higher-NXB X-ray instruments.
\item Altough the coverage of sky with Xtend is greatly limited compared with all-sky monitors such as \textit{MAXI}, its sensitivity is $\sim 4$ orders of magnitude higher than \textit{MAXI}, and the FOV is significantly wider than most other modern X-ray imaging instruments such as \textit{Chandra}/ACIS and \textit{XMM-Newton}/EPIC. 
\end{enumerate}
\end{enumerate}

 Then, the following requirements  were  set for Xtend to satisfy the desired features. 
\begin{itemize}
\item As a focal-plane X-ray sensor, a back-illuminated CCD chip is superior, considering the overall balance, notably its high  tolerance to micrometeoroids and orbital debris.
\item The FOV must be larger than $22' \times 22'$ for  feature items 1-(a, b, d) and 2-(a, b) in the list above. 
\item The effective area must be larger than $300$~cm$^{2}$ at 1.5~keV and larger than $270$~cm$^{2}$ at 6~keV for feature items 1-(a, b, c, d) and 2-(a, b). 
\item The pixel size must be below $100~\mu$m, with which the PSF core of the XMA (\cite{Tamura2022}) can be spatially resolved (1-(e)). 
\item The detector must be sensitive for the energy range  from 0.4 to 13~keV to cover almost the same range as Resolve (1-(a, b, c, d) and 2-(b)). 
\item The energy resolution at 6~keV must be  better than $200$~eV  in\ Full Width Half Maximum (FWHM) at the Beginning-Of-Life (BOL) and $250$~eV (FWHM) at the End-Of-Life (EOL) (1-(d) and 2-(a)). 
\item The optical blocking performance (light reduction)   must be $< 10^{-13}$ along the X-ray light path including the reduction by (the structure of) the satellite itself of $10^{-6}$ (1-(a, b, c, d) and 2-(a, b)).  
\item The NXB level must be below $1\times 10^{-6}$~counts~keV$^{-1}$~s$^{-1}$~arcmin$^{-2}$~cm$^{-1}$ in 5--10~keV (1-(a, d) and 2-(a, b)). 
\end{itemize} 
%要求仕様書の付録を確認して、矛盾がなければ OK。こちらの方が一般的に書かれている。

%%%%%%%%%%%Table 1%%%%%%%%%%                                                                                                                              
\renewcommand{\arraystretch}{1}
\begin{table*}[t]
\caption{Summary of the design changes from the \textit{Hitomi}/SXI to \textit{XRISM}/SXI.}
 \begin{center}
 \begin{tabular}{cccc}
\hline\hline
Component & Design change & Reason for the change & Section
\\\hline
%SXI-S-BDY & Route and length of the vent pipe & To prevent stray light and improve workability &  \S3.1.1 \\
%		& Extension of the MLI slit & To improve workability on the side box MLI & \S3.1.1 \\[1.5ex]
CCD chips %& Focal point & Due to the change of the XMA set position & - \\ 
 		& Thickness of the OBL & To prevent light leakage through pinholes & \S3.1.2\\
		 & Additional Al layer & To prevent light leakage via the physical edges & \S3.1.2 \\
		 & Notch implant & To suppress the charge transfer inefficiency & \S3.1.2 \\[1.5ex]
%SXI-S-FE & Additional circuit & To prevent the loop current anomaly & \S3.1.4 \\[1.5ex]
SXI-S-1ST & Reduction of one of the two 1STs & Based on the \textit{Hitomi}/SXI  performance & \S3.1.5 \\[1.5ex]
SXI-PE & Reduction of clocking modes &  Based on the  necessity in orbit  & \S3.2.2 %\\[1.5ex]
%System  	& Reduction of X-MDE &   & - \\
%		  &  Reduction of a lashing point on the SXI & To improve workability & - \\
%		  & Fill of the hole for \textit{Hitomi}/HXI & To prevent stray light via the hole for HXI & - \\
\\\hline\hline

\end{tabular}
\end{center}
\label{tab1}
\end{table*}
%%%%%%%%%%%Table 1%%%%%%%%%%%% 

 We found that the best solution that satisfies all these requirements  was  a combination of the XMA and the CCD-chip model  ``PchNeXT4'' fabricated by Hamamatsu Photonics K.K  (\cite{Matsuura2006}; \cite{Ozawa2006}; \cite{Takagi2006}; \cite{Ueda2011}; \cite{Ueda2013}). 
The PchNeXT4 chip  is a P-channel Back-Illuminated (BI)-type CCD with a 200-$\mu$m-thick depletion layer.  Four chips of the type were employed in the X-ray CCD camera onboard \textit{Hitomi} \citep{Tanaka2018}, which is also called the SXI, and their performances have  already  been studied well on  ground and in orbit \citep{Nakajima2018}. 
 For the \textit{XRISM}/Xtend/SXI, we  developed a variant of PchNeXT4, applying some improvements that we considered desirable after examining  the in-orbit performance of the original PchNeXT4 chips, and named the new CCD chips ``PchNeXT4A.''
  The next section \S3 provides a detailed description of the improvements that we made. 
   For the \textit{XRISM}/Xtend/SXI, we arranged  four PchNeXT4A chips in a $2 \times 2$ array.  
As shown from the next section, the \textit{XRISM}/Xtend/SXI can be considered one of the best X-ray CCD cameras ever installed on X-ray astronomical satellites, offering a relatively large FOV with a stable NXB level and achieving the largest grasp at 7~keV  among X-ray CCD cameras onboard large observatories when combined with the XMA (see Uchida et al. in prep for the in-orbit performance of Xtend).
 Hereafter, we  refer to the SXI of the Xtend system onboard \textit{XRISM} as the \textit{XRISM}/SXI or just SXI,  and the SXI onboard \textit{Hitomi} as the \textit{Hitomi}/SXI, to distinguish them clearly.

%============S3============
\section{ Design of the \textit{XRISM}/SXI}
%============S3============
%==========================
%\subsection{SXI system}
%==========================

%\begin{verbatim}
\begin{figure*}
 \begin{center}
  \includegraphics[width=140mm]{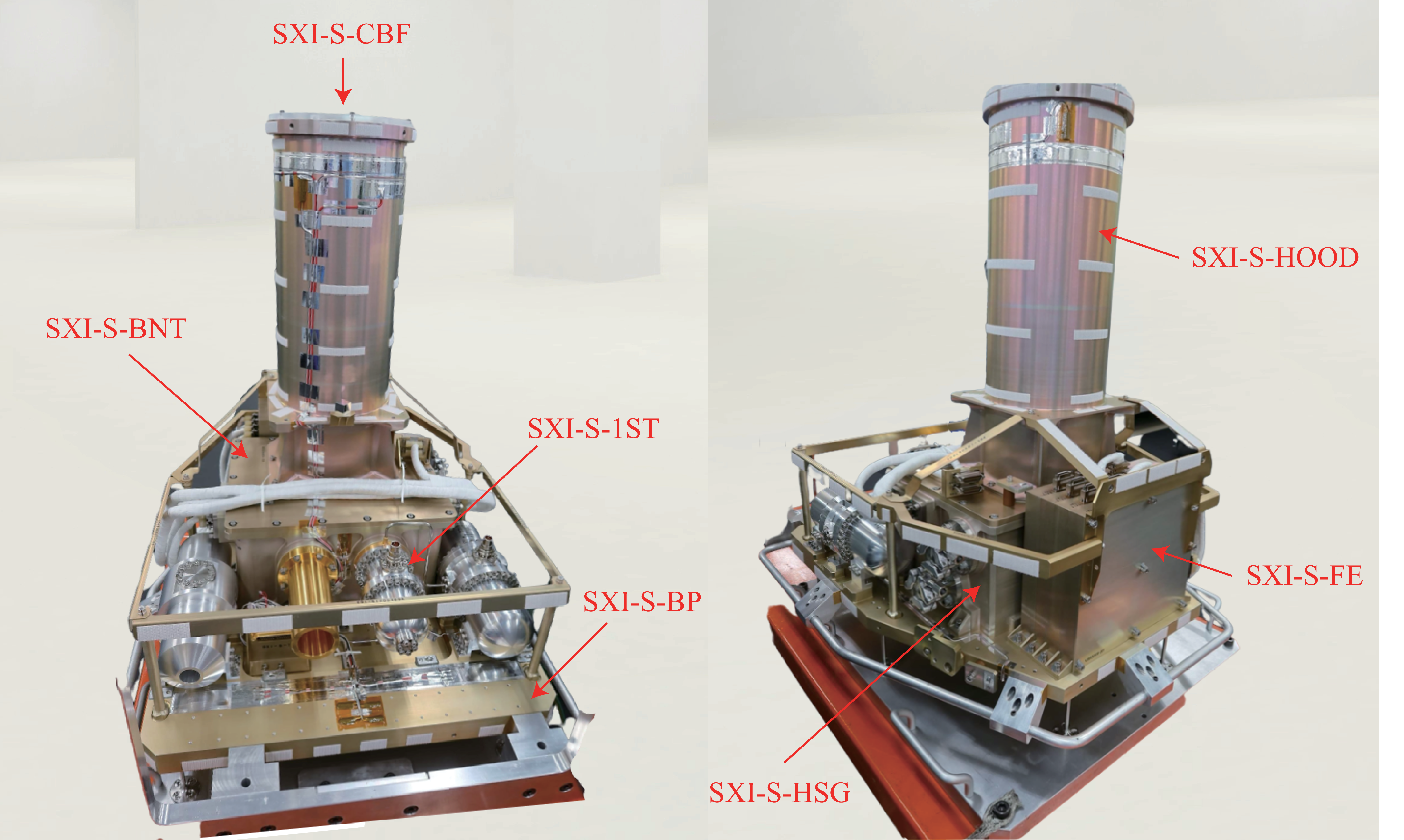}
 \end{center}
 \caption{Photographs of the (left panel) front  and (right) back  sides of SXI-S (credit: JAXA). {Alt text: Two photographs.}}\label{fig2}
\end{figure*}
%\end{verbatim}

Figure~\ref{fig1} shows a block diagram of the \textit{XRISM}/SXI. 
  The \textit{XRISM}/SXI is composed of four parts of the sensor part (SXI-S), the pixel processing electronics (SXI-PE), the digital electronics (SXI-DE), and the cooler driver (SXI-CD), in the same way as in the design of and naming convention for the \textit{Hitomi}/SXI \citep{Tanaka2018}. 
In this section, we present a brief summary of each part   (refer to the design description of the \textit{Hitomi}/SXI  presented in \citet{Tanaka2018} for detail),    except for the changes that we made from the \textit{Hitomi}/SXI, for which we give a full description here.  Table~\ref{tab1}   summarizes the changes in one sentence for each. 
%All components described in this section  are  manufactured and assembled by Mitsubishi Heavy Industry, Ltd. unless otherwise noted (e.g., the CCD chips were fabricated by Hamamatsu Photonics K.K, as mentioned in \S2). 
Most of the components described in this section  are  manufactured by  Mitsubishi Heavy Industry, Ltd., while the CCD chips were fabricated by Hamamatsu Photonics K.K, as mentioned in \S2, and the others were provided by JAXA. Mitsubishi Heavy Industry, Ltd. assembled them into the SXI. 

%==========================
\subsection{Sensors and front-end components: SXI-S}
%==========================

Figure~\ref{fig2} shows an overall picture of SXI-S, which 
 is a sensor part at the focal plane of the XMA,  located on the base plate of the spacecraft.
SXI-S consists of the camera body (SXI-S-BDY; \S\ref{camera}), single-stage Stirling cooler (SXI-S-1ST; \S\ref{1st}), video boards (\S\ref{video}), and four driver boards (SXI-S-FE; \S\ref{fe}).
The CCD driving clock patterns, operating commands, and heater control commands are provided to SXI-S from SXI-PE, whereas the driving power of SXI-S-1ST is independently input from SXI-CD. 
CCD image data and House Keeping (HK) data are digitized in the video and driver boards of SXI-S and then transferred to SXI-PE.

%\begin{verbatim}
\begin{figure*}
 \begin{center}
  \includegraphics[width=120mm]{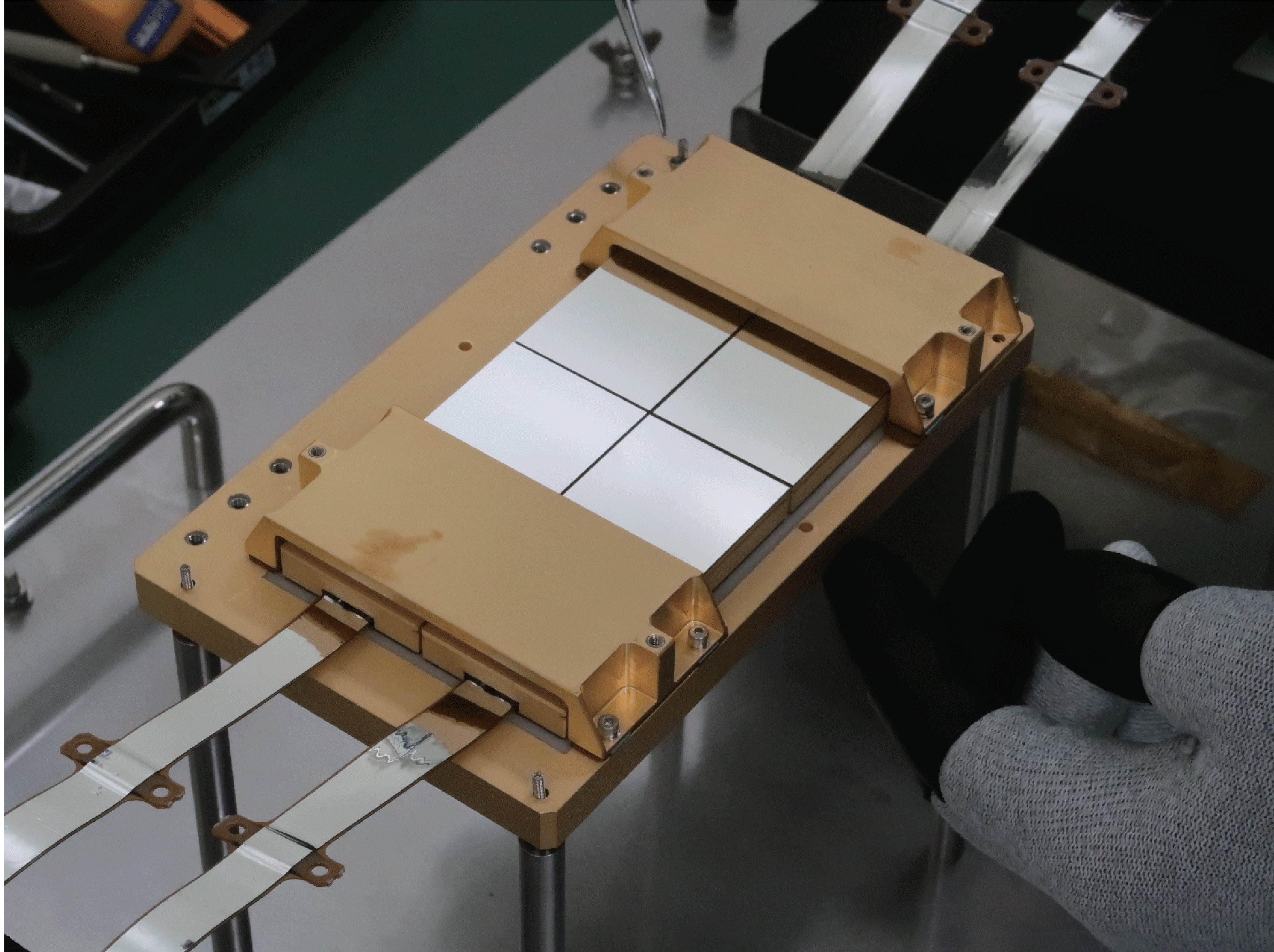}
 \end{center}
 \caption{Photograph of the flight-model CCD chips mounted on the cold plate. The four CCD chips are mounted on the cold plate supported by jigs. The frame-store regions of the CCD sensors are shielded by the frame-store cover, and a Flexible Printed Circuit (FPC) extends from each of the four CCD sensors. {Alt text: One photograph.}}\label{fig3}
\end{figure*}
%\end{verbatim}

%==========================
\subsubsection{Camera body: SXI-S-BDY}
\label{camera}
%==========================

The camera body, SXI-S-BDY, is made of Al alloy (Figure~\ref{fig2}). In this subsection, we describe the relative positions of the components of SXI-S-BDY in the orientation in which the SXI observes X-rays from sources located directly \textit{above} SXI-S-BDY as in Figure~\ref{fig2}. 
The main components of SXI-S-BDY are a focal plane assembly (SXI-S-FPA), housing (SXI-S-HSG), bonnet (SXI-S-BNT), hood (SXI-S-HOOD), contamination blocking filter (SXI-S-CBF), vent pipes (SXI-S-VP), and base plate (SXI-S-BP).

SXI-S-BNT is a lid set at the top of SXI-S-HSG. 
SXI-S-HOOD is a baffle set above SXI-BNT to  block stray X-rays without sacrificing a wide FOV. 
SXI-S-CBF is  installed at the top of SXI-S-HOOD and spatially separates inside and outside of SXI-S-BDY to block  contamination. 
The design of SXI-S-CBF is the same as that of the \textit{Hitomi}/SXI, i.e., a 200-nm polyamide layer is sandwiched by 80-nm and 40-nm Al layers (n.b.,\ the presented thicknesses are design values).   
SXI-S-FPA consists of four CCD chips of PchNeXT4A (see \S3.1.2), a cold plate with heaters, video boards (see \S3.1.3), and mechanical support  structure inside SXI-S-HSG. 
SXI-S-VP is composed of two vent pipes for venting remaining gases in SXI-S-BDY outside the spacecraft. 
%for the remaining gas inside SXI-S-BDY to vent outside of the spacecraft. 
SXI-S-BP is positioned approximately ten  centimeters above the base plate of the spacecraft using Ti alloy shafts, preventing distortion due to thermal expansion and minimizing thermal interaction between SXI-S and the spacecraft through conduction. 
Multi-Layer Insulators (MLIs) cover SXI-S-BDY, which suppresses  heat exchange by radiation.
The heat produced in SXI-S is carried to the radiators through heat pipes connected to the SXI-S-BP.
For  a countermeasure to  molecular contamination inside SXI-S-BDY, we baked every component at $\sim 100^{\circ}$C before the launch until the amount of outgas became lower than the detection limit. 
%As a result, we confirmed that the amount of outgas can be suppressed to $< 4~\mu$g/cm$^2$ after 3 years in-orbit operation.
With this baking process, the amount of molecular contamination absorbed onto the surfaces of the CCDs from SXI-S-BDY should remain $< 4~\mu$g/cm$^2$ after 3 years in-orbit operation (i.e., untill the EOL) according to our model estimate.

Two calibration sources of $^{55}$Fe radioisotopes are placed at the surface of SXI-S-BNT above the CCD chips to measure the imaging and spectroscopic performance of the SXI in orbit.  
 X-ray photons from the calibration sources are irradiated to the corners of the four CCD chips, and each irradiated  region is a  circular quadrant with a radius  smaller than $4.5$~mm. 
To determine the energy gain at 5.9~keV within $0.1$\% accuracy, the count rate of $0.4$~counts/sec/CCD is required. 
Considering a half-life of $^{55}$Fe of 2.73~years, the radiation intensity at 5.9~keV of the calibration sources is required to be more than $0.9$~counts/sec/CCD at the BOL in order that the intensity of more than $0.4$~counts/sec/CCD is maintained throughout the effective mission lifetime of three years. 

%\begin{verbatim}
\begin{figure*}
 \begin{center}
  \includegraphics[width=150mm]{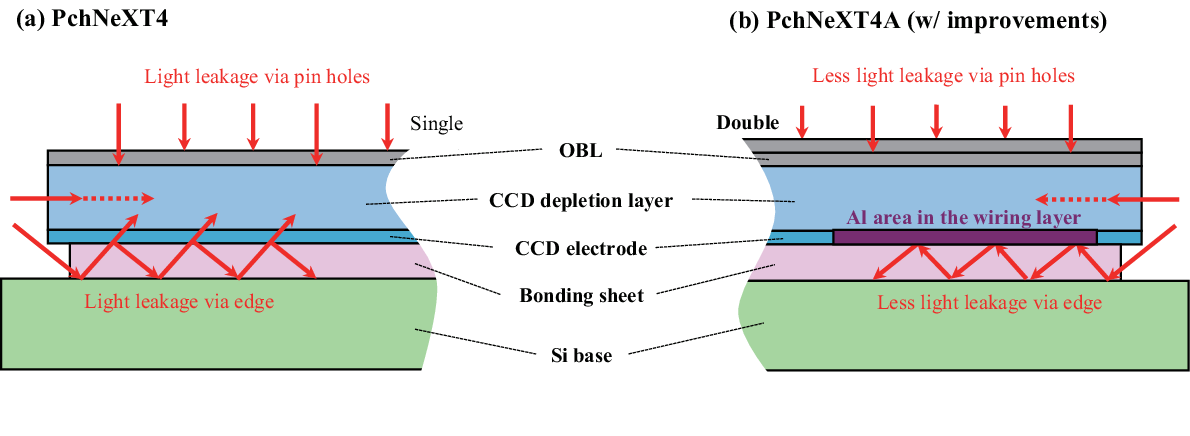}
 \end{center}
 \caption{Schematic  illustrations of (left) PchNeXT4  and (right) PchNeXT4a. PchNeXT4A is improved from PchNeXT4 with  a doubled OBL and an additional aluminum area between the depletion layer and the bounding sheet (see \S3 and Table~\ref{tab1} for detail). {Alt text: Two schematic illustrations.}}\label{fig4}
\end{figure*}
%\end{verbatim}

%==========================
\subsubsection{CCD chips (PchNeXT4A)}
\label{ccd}
%==========================

%%%%%%%%%%%Table 1%%%%%%%%%%                                                                                                                              
\renewcommand{\arraystretch}{1}
\begin{table*}[t]
\caption{Specifications of the CCD chip (PchNeXT4A) for  \textit{XRISM}/SXI.}
 \begin{center}
 \begin{tabular}{cc}
\hline\hline
Architecture & Frame Transfer 
\\\hline
Channel type & P-channel
\\
Chip size & 31.220~mm (horizontal) $\times$ 57.525~mm (vertical)
\\
Imaging area size & 30.720~mm $\times$ 30.720~mm 
\\
Pixel format (physical/logical) & 1280 $\times$ 1280 /  640 $\times$ 640 
\\ 
Pixel size (physical/logical) & 24~$\mu$m $\times$ 24~$\mu$m /  48~$\mu$m $\times$ 48~$\mu$m
\\ 
Depletion layer thickness & 200~$\mu$m (design value)
\\ 
Incident surface coating & 100~nm + 100~nm thick Al (design value)
\\
Readout nodes (equipped/used) & 4 / 2 
\\
Output & 1-stage MOSFET source follower 
\\
Charge conversion efficiency & $5~\mu$V$/e^{-}$ (design value)
\\
Readout noise (whole SXI system) & 6--7~$e^{-}$ (rms)
\\\hline\hline

\end{tabular}
\end{center}
\label{tab2}
\end{table*}
%%%%%%%%%%%Table 1%%%%%%%%%%%% 

The \textit{XRISM}/SXI employs  P-channel Back-Illuminated (BI)-type CCDs with a 200~$\mu$m-thick depletion layer, named PchNeXT4A (see \S2). 
Figure~\ref{fig3} shows a photograph of the four flight-model CCD chips mounted on the cold plate, prior to their installation inside SXI-S-HSG.
 Owing to  its thick depletion layer, they  achieve a higher Quantum Efficiency (QE) and lower NXB level at around 6~keV than those of the BI-type CCD camera of \textit{Suzaku}/XIS with a 40~$\mu$m-thick depletion layer \citep{Koyama2007}.  
Their charge-transfer architecture is the frame transfer type, and  accordingly, each of the CCD chips has an imaging area and a frame-store area. 
The imaging area has a size of $30.720~\textrm{mm} \times 30.720~\textrm{mm}$ and $1280 \times 1280~\textrm{physical~pixels}$. 
In default,  $2 \times 2$ on-chip binning is performed first prior to any following processing; as a result, a derived image  has $640 \times 640~\textrm{logical~pixels}$ with a pixel size of $48~\mu\textrm{m} \times 48~\mu\textrm{m}$ each. 
The imaging area and frame-store area are horizontally divided into two segments along the central line of the device, aligned with the columns.
The signals from each segment are read out using one of the two readout nodes installed at both ends of the bottom of the frame-store area of each segment.
The CCD operation temperature is $-110$ or $-120^{\circ}$C, either of which is sufficiently low to suppress dark currents to the acceptable level. 
Table~\ref{tab2}  summarizes the specifications of the PchNeXT4A chip. 

 \citet{Nakajima2018} reported that the most serious concern in the PchNeXT4 chips of \textit{Hitomi}/SXI was optical light leakage.  
They identified two main paths for leakage:    through many pinholes in the Optical Blocking Layer (OBL) and  through the physical edges of the chips.
 Optical leakage caused a significant amount of pseudo-events with the \textit{Hitomi}/SXI in orbit, especially when the base plate of the \textit{Hitomi} spacecraft was pointed to the day Earth,  where optical/infrared photons passed  through a hole for Extended Optical Bench (EOB) for the Hard X-ray Imager (HXI) and reached the detector surface. The leakage ultimately resulted in a significant reduction in the observation efficiency.
To resolve this, we filled the hole of the spacecraft base plate that had been necessary for the HXI in \textit{Hitomi}; this was possible because \textit{XRISM} lacks the HXI unlike \textit{Hitomi}, even though the basic design of \textit{XRISM} follows that of \textit{Hitomi}, including the base plate. 
Additionally, we directly applied two improvements to the PchNeXT4A chips for the \textit{XRISM}/SXI from PchNeXT4 for the \textit{Hitomi}/SXI to suppress the optical/infrared light leakage.    
First, we  doubled the number of the OBL to reduce the number of pinholes of the X-ray incident surface. As a result,  the total thickness of the aluminum layer  was increased from $100~{\rm nm}$ to $200~{\rm nm}$.
Second, we  inserted an additional aluminum layer between the bonding sheet and the depletion layer to prevent light leakage from the physical edges of the chips. Figure~\ref{fig4} illustrates the cross-sectional views of PchNeXT4 and PchNeXT4A, highlighting the improved points. 
Consequently, the optical/infrared blocking performance of the PchNeXT4A chips of the \textit{XRISM}/SXI was confirmed to be significantly improved from the PchNeXT4 chips of the \textit{Hitomi}/SXI, as reported by \citet{Uchida2020}.    

%\begin{verbatim}
\begin{figure}
 \begin{center}
  \includegraphics[width=60mm]{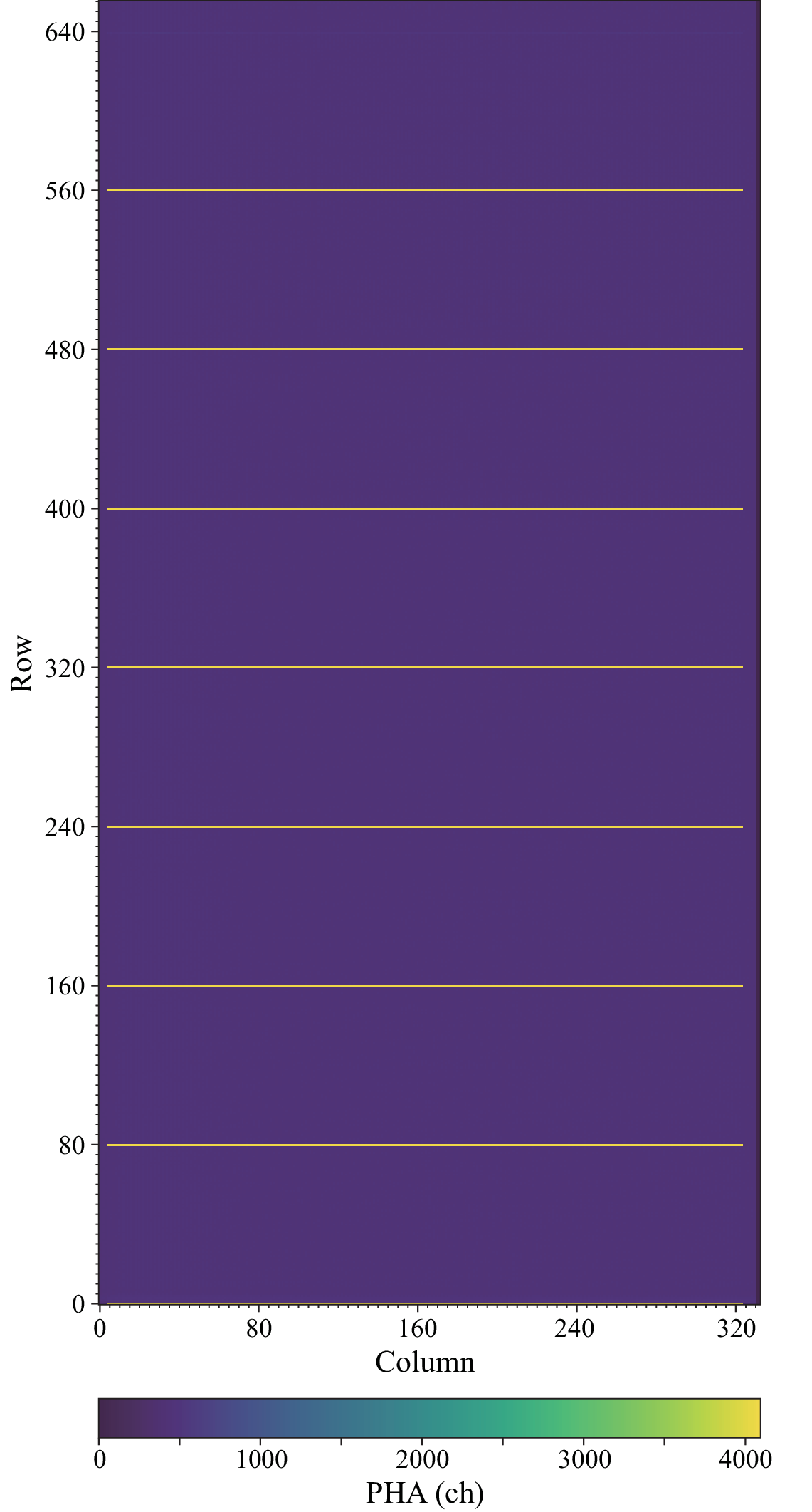}
 \end{center}
 \caption{Frame image obtained from one segment of a CCD chip of the \textit{XRISM}/SXI, with the color scale representing pulse heights (ch). Charge injection (CI) lines saturated at 4095~ch are located every 80 rows. {Alt text: One frame image}}
 \label{fig:frame}
\end{figure}
%\end{verbatim}

We applied another improvement, related to the expected  degradation of the energy resolution over three years (planned mission lifetime) in orbit, to the PchNeXT4A chips for the \textit{XRISM}/SXI. 
The degradation is partly  the result of the expected increase  in the CTI caused by accumulated charge traps in transfer paths in the CCD chips. In addition to the existing charge traps at the time of the launch, it is well known that the number of traps increases in orbit over time due to radiation damage by cosmic rays, which increases the CTI. 
 Since the number of charge transfers  varies, depending on  the X-ray photon incident positions in the imaging area, the  obtained pulse height amplitude for an event of a certain energy of incident photon varies according to the incident position, and its uncertainty increases  as the CTI increases over time, which results in  degradation of the energy resolution.    
In order to minimize the CTI increase in orbit, we implanted a narrow notch  in the charge transfer path to confine a charge packet to a limited fraction of the pixel width ($24~\mu$m) in the PchNeXT4A chips.  
In fact, \citet{Kanemaru2019} reported that CCD chips with notch implants which are equivalent to the flight-model CCDs showed a $\sim 3$ times higher radiation hardness than those without them. Thus, we expect a similar improvement in the radiation hardness for PchNeXT4A with our notch implanting. 

Furthermore, the charge injection (CI) technique is routinely applied to the \textit{XRISM}/SXI during the operation in orbit to reduce the CTI  in the same way as with the \textit{Suzaku}/XIS (e.g., \cite{Prigozhin2008}; \cite{Nakajima2008}; \cite{Uchiyama2009}; \cite{Ozawa2009}) and \textit{Hitomi}/SXI (\cite{Nobukawa2014}; \cite{Tanaka2018}). The concept is that once CI is conducted, charge traps are filled with artificial charges before charges generated by X-ray events pass the points in order that the potential existent charge traps no longer trap charges from X-ray events. 
With the \textit{XRISM}/SXI, CI is conducted once per 80 vertical logical-pixel transfers. 
In each CI, horizontally-aligned artificial charges are injected into the imaging area from the serial register at the top. 
Figure~\ref{fig:frame} shows an example of a frame image obtained by one segment of a CCD chip, where CI lines with pulse heights saturated at 4095~ch are confirmed every 80~rows.

%==========================
\subsubsection{Video board}
\label{video}
%==========================

 SXI-S-HSG accommodates, in addition to four CCD chips, two video boards, each of which is  responsible for processing the front-end signal  of two CCD chips  (see Figure~\ref{fig1}). 
 The main part of the video board consists of two Video Application Specific Integrated Circuits (Video ASICs) named MND02 \citep{Nakajima2011} and a Video Field-Programmable Gate Array (Video FPGA) with the model number RTAX2000. 

The MND02 chips conduct three functions;  firstly, it works as a preamplifier of analog signals by a factor of 0.63 to 10 (a factor of 10 is employed in the normal mode),  secondly, it performs an analog offset function, and lastly,  it does  an Analog to Digital (AD) conversion of CCD signals employing the $\Delta$-$\Sigma$ modulation. 
In the AD conversion, two modulators   are implemented, which work for either of the even and odd columns, to speed up signal processing. 
 We adopted the spare MND02 chips  for the \textit{Hitomi}/SXI \citep{Nakajima2013} as the flight models of the \textit{XRISM}/SXI,  the performance and quality of  which  had been verified to be sufficient for  use in orbit. 

The Video FPGA performs digital filter processing on the $\Delta$-$\Sigma$ modulation stream data from the ASICs and converts them into pulse-height data. 
After the AD conversion, the data of digitized frames are multiplexed and transferred to SXI-PE for subsequent processing. 
The Video FPGA also controls external interfaces connected to SXI-PE and SXI-S-FE. 

%==========================
\subsubsection{Front-end electronics: SXI-S-FE}
\label{fe}
%==========================

SXI-S-FE has four driver boards, one for one CCD chip,  mounted on SXI-S-BP (Figure~\ref{fig1}).
 The driver board has  four main functions. 
First,  it generates bias and clock voltages and supplies them to the CCD chips via the video boards. 
 Second,  it measures   the CCD temperature and outputs the AD-converted value.  
Third, it obtains the HK data of the video boards and driver boards, and conducts AD conversion.
 And fourth,  it supplies the power to the heaters on the cold plate, on which the CCD chips are mounted. 
 In pre-flight tests of the \textit{Hitomi}/SXI and  monitoring in orbit, we observed  occasional  events of current anomaly in the driver board  when  the driver was powered up (\cite{Hayashida2018}). 
We  added a countermeasure circuit to SXI-S-FE of the \textit{XRISM}/SXI  to suppress the current-anomaly events.  
%We classified these events into several patterns and performed intensive investigations to design the \textit{XRISM}/SXI. 
%As a result, we identified that they are due to the initialization error by the loop current or analog switch malfunction of the AD converters. 
%In addition, we found that the current anomaly events did not cause the hardware damage, and the driver boards recovered their normal state after restart.  
%Therefore, 
%We newly added a countermeasure circuilt for the loop current abnormality to SXI-S-FE of \textit{XRISM}/SXI. 
%In the case that the current anomaly events occur, we just restart the driver boards in orbit.  

%==========================
\subsubsection{Single stage Stirling cooler: SXI-S-1ST}
\label{1st}
%==========================

\textit{XRISM}/SXI employs  a single Stirling cooler manufactured by Sumitomo Heavy Industries, Ltd,  named SXI-S-1ST,  the model of which is basically the same as that for the \textit{Hitomi}/SXI. 
A difference is that we installed only one cooler to the \textit{XRISM}/SXI, as opposed to two (equivalent ones) for the \textit{Hitomi}/SXI  \citep{Tanaka2018}, because only one is sufficient to cool the CCD temperature to $-120~^{\circ}$C. The \textit{Hitomi}/SXI had two for redundancy.   
  We  instead installed a dummy cooler   in the vacant space.  

%
%%%%%%%%%%%%Table 1%%%%%%%%%%                                                                                                                              
%\renew{\arraystretch}{1}
%\begin{table*}[t]
%\caption{The clocking modes of \textit{XRISM}/SXI. }
% \begin{center}
% \begin{tabular}{ccccc}
%\hline\hline
%Mode name & Logical pixels (H $\times$ V) & Exposure time (sec) & Exposure per frame & User Support \\\hline
%Full window + no burst & $640 \times 640$ & 3.9631 & 1 & Yes \\
%Full window + 0.1~sec burst & $640 \times 640$ & 0.0606 & 1 & No \\ 
%$1/8$ window + no burst & $640 \times 80$ & 0.4631 & 8 & Yes \\ 
%$1/8$ window + 0.1~sec burst & $640 \times 80$ & 0.0606 & 8 & Yes \\ 
%\hline\hline
%
%\end{tabular}
%\end{center}
%\label{tab2}
%\end{table*}
%%%%%%%%%%%%Table 1%%%%%%%%%%%% 

%==========================
\subsection{Pixel process electronics: SXI-PE}
%==========================

\subsubsection{Components and functions of SXI-PE}

SXI-PE has one Power Supply Unit (PSU) board and two identical Mission I/O (MIO) boards (Figure~\ref{fig1}). 
The PSU board supplies the DC power to the two MIO boards, the four FE boards, and Video FPGAs in SXI-S from the bus power of the spacecraft. 
Each MIO board is comprised of two FPGAs, SpaceWire FPGA, User FPGA, and Synchronous Dynamic Random Access Memory (SDRAM). 
The SpaceWire FPGA controls communications over the SpaceWire. 
The User FPGA manages functions based on user-specific logic, such as the clock-pattern generation (sequencer) for the CCD driving, 
%the $\Delta$-$\Sigma$ AD conversion, 
 processing  the CCD data,  collecting  the HK data from SXI-S-FE, and  controlling  Digital-Analog Conversions (DACs) on SXI-S-FE. 
One MIO board provides   identical driving patterns to two of the four CCD chips and transmits their data to SXI-DE. 
To avoid signal interference among the CCD chips, one of the two MIO boards provides a master clock, and both MIO boards share it. 
The timing signals follow the 10~MHz clock, and thus, the logical levels of individual signals can be assigned in every $0.1~\mu$sec. 
Another important function of SXI-PE is the on-board data processing of signals from the SXI-S. 
This function is summarized in \S4. 

\subsubsection{CCD clocking modes}

%%%%%%%%%%%Table 1%%%%%%%%%%                                                                                                                              
\renewcommand{\arraystretch}{1}
\begin{table*}[t]
\caption{Clocking modes of \textit{XRISM}/SXI. }
 \begin{center}
 \begin{tabular}{ccccc}
\hline\hline
Mode name & Logical pixels (H $\times$ V) & Exposure time (sec) & Exposure per frame & Available for guest observers? \\\hline
Full window + no burst & $640 \times 640$ & 3.9631 & 1 & Yes \\
$1/8$ window + no burst & $640 \times 80$ & 0.4631 & 8 & Yes \\ 
$1/8$ window + burst & $640 \times 80$ & 0.0620 & 8 & Yes \\ 
Full window + burst & $640 \times 640$ & 0.0606 & 1 & No \\ 
\hline\hline

\end{tabular}
\end{center}
\label{tab3}
\end{table*}
%%%%%%%%%%%Table 1%%%%%%%%%%%% 

The timing of CCD-driving voltage clocks and $\Delta$-$\Sigma$ AD conversions are determined by the microcode programs loaded to SXI-PE,  in the same way as used in the \textit{Suzaku}/XIS and \textit{Hitomi}/SXI. 
The unit times of charge transfer are 14.4~$\mu$sec (69.444~kHz) in the horizontal direction (per logical pixel) and 28.8~$\mu$sec (34.722~kHz) in the vertical direction (per physical pixel). 
Various clocking modes are  available according to the numbers and patterns of horizontal and vertical transfers.  
Four of the  five clocking modes  prepared  for the \textit{Hitomi}/SXI \citep{Tanaka2018} are employed for in-orbit observations of the \textit{XRISM}/SXI, where the excluded one  is the full-window + 2~sec burst mode.  

The four clocking modes of \textit{XRISM}/SXI are  described in the following list and in Table~\ref{tab3}. 
\begin{itemize}
\item The ``full window + no burst'' mode reads out the entire imaging area with an exposure time of 3.96~seconds per image. 
This mode is used to read out all X-ray events with no  spatial  or temporal constraints and is called the normal mode. 

\item The ``$1/8$ window + no burst'' mode reads out one-eighth of the imaging area 8 times in 4~seconds with an exposure time of 0.46~seconds  per readout.  
This mode is effective in reducing pile-up effects for bright sources, for which  most X-ray events would need to be discarded in the normal mode to reduce the pile-up effects to a similar level. 

\item The ``$1/8$ window + burst'' mode is basically the same as the ``$1/8$ window + no burst'' mode  except for the exposure time per image  of 0.06~seconds, reduced from 0.46~seconds; i.e., 
this mode reads out signals only  after the first 0.06~second exposure in each frame and reads them blankly in the rest $0.44$~seconds. 
This mode  most suppresses the pile-up effects among the four clocking modes and is suitable for observations of the brightest sources. 

\item The ``full window + burst'' mode  reads out the entire imaging area at 4-sec intervals  but limits the exposure time per image to $\sim 0.06$~sec in a similar way as in the ``$1/8$ window + burst'' mode. 
Whereas this  mode highly suppresses pile-up effects caused by a large number of X-ray events when observing bright sources, the  dead time is 3.94~seconds per frame, which is over eight times longer than in the ``$1/8$ window + burst'' mode. 
This mode  is meant for  diagnostic purposes only and is not  available for general users of \textit{XRISM}. 
\end{itemize}

%==========================
\subsection{Digital electronics: SXI-DE}
%==========================

SXI-DE is composed of one PSU board and one Central Processing Unit (CPU) board called Space Card\ (Figure~\ref{fig1}). 
The PSU board supplies  DC power to the Space Card from the spacecraft bus power. 
The Space Card acquires event candidates and frame data extracted by SXI-PE, and performs further data processing, such as  event selection. 
The on-board data processing by SXI-DE is  described in \S4. 
After the data processing, SXI-DE converts the event and frame data into telemetry. 

Another function of SXI-DE is to serve as the interface between the system of the SXI and the bus system of the spacecraft. 
SXI-DE receives commands provided by the Satellite Management Unit (SMU) in the bus system of the spacecraft. 
 Conversely, SXI-DE  transmits the telemetry of events and the data of HK values to the SMU and the Data Recorder (DR). 
In addition, SXI-DE oversees the entire SXI system, excluding SXI-S-1ST which is managed by SXI-CD.  

%==========================
\subsection{Cooler driver: SXI-CD}
%==========================

Power to operate SXI-S-1ST is supplied by SXI-CD,  manufactured by Sumitomo Heavy Industries, Ltd. 
SXI-CD receives the satellite bus power and generates a sine-wave current to drive the refrigerator  compressor and active balancer of  SXI-S-1ST. 
SXI-CD controls the cooler according to the commands received from the SMU.
%The cooler controlling is performed, according to the commands received from the SMU.
Furthermore, SXI-CD monitors the driving voltage, current, and power of the refrigerator compressor and active balancer  and outputs their values as telemetry. 
In the SXI, SXI-DE regulates the CCD temperature  at a target value   with the Proportional-Integral-Derivative (PID) control, which increases or decreases the heater current based on its the need. 
Therefore, the SXI-CD does not need or have a temperature sensor for feedback control. 

%===========S3===============
\section{Data Processing}
%============S3==============

%\begin{verbatim}
%\begin{figure*}
% \begin{center}
%  \includegraphics[width=160mm]{figure/fig5.eps}
% \end{center}
% \caption{Pulse heights of grade 0 events in an energy band containing Mn-K$\alpha$ and Mn-K$\beta$ lines, plotted against row number (half of the transfer count due to on-chip binning), before and after CTI correction, in the left and middle panel, respectively. The right panel shows spectra before and after CTI corrections. These data were obtained in the SXI thermal vacuum test shown in \S4.4. }\label{fig5}
%\end{figure*}
%\end{verbatim}

The events detected by the CCD chips, both X-ray and non-X-ray events, are
%X-ray events detected by the CCD chips are 
processed and screened on board by SXI-PE and SXI-DE.
After the screened event data are donwlinked, we correct the pulse height amplitudes on  ground in four steps. 
This section describes the on-board and on-ground data processing and corrections. 

%------------------------------------------------------------
\subsection{On-board processing}
%------------------------------------------------------------
%------------------------------------------------------------
\subsubsection{Event candidate extraction in SXI-PE}
%------------------------------------------------------------

The CCD analog output is first AD-converted by the video board and  then transmitted to  SXI-PE, where the data are pre-processed in the UserFPGA.
SXI-PE selects designated ASIC channels  according to the registered settings and calculates the average values of signals  from the  readout nodes. 
 Then, SXI-PE estimates  the dark level of each pixel, which  temporally fluctuates pixel by pixel,  and updates it at every frame. 
Refer to \citet{Tanaka2018} for details on the pixel-by-pixel estimation and updates of the dark level as well as the identification of hot pixels.
%Pixels with dark levels exceeding a preset threshold are flagged as hot pixels. 
Subsequently, SXI-PE subtracts the dark level from the raw frame data, and the UserFPGA program identifies X-ray event candidates using the predefined thresholds and relations of pulse-height values in multiple pixels. 
The raw frame data, dark image (Dark image), pixel-value image (Frame image), event-candidate list, and hot-pixel list are stored in SDRAM.
Among them, the Frame image, event candidate list, and hot pixel list are transmitted to SXI-DE for further data processing. 
Although the event-candidate extraction in SXI-PE is simple,  as described above, it improves the efficiency of subsequent processing in SXI-DE. 

%\begin{verbatim}
\begin{figure*}
 \begin{center}
  \includegraphics[width=160mm]{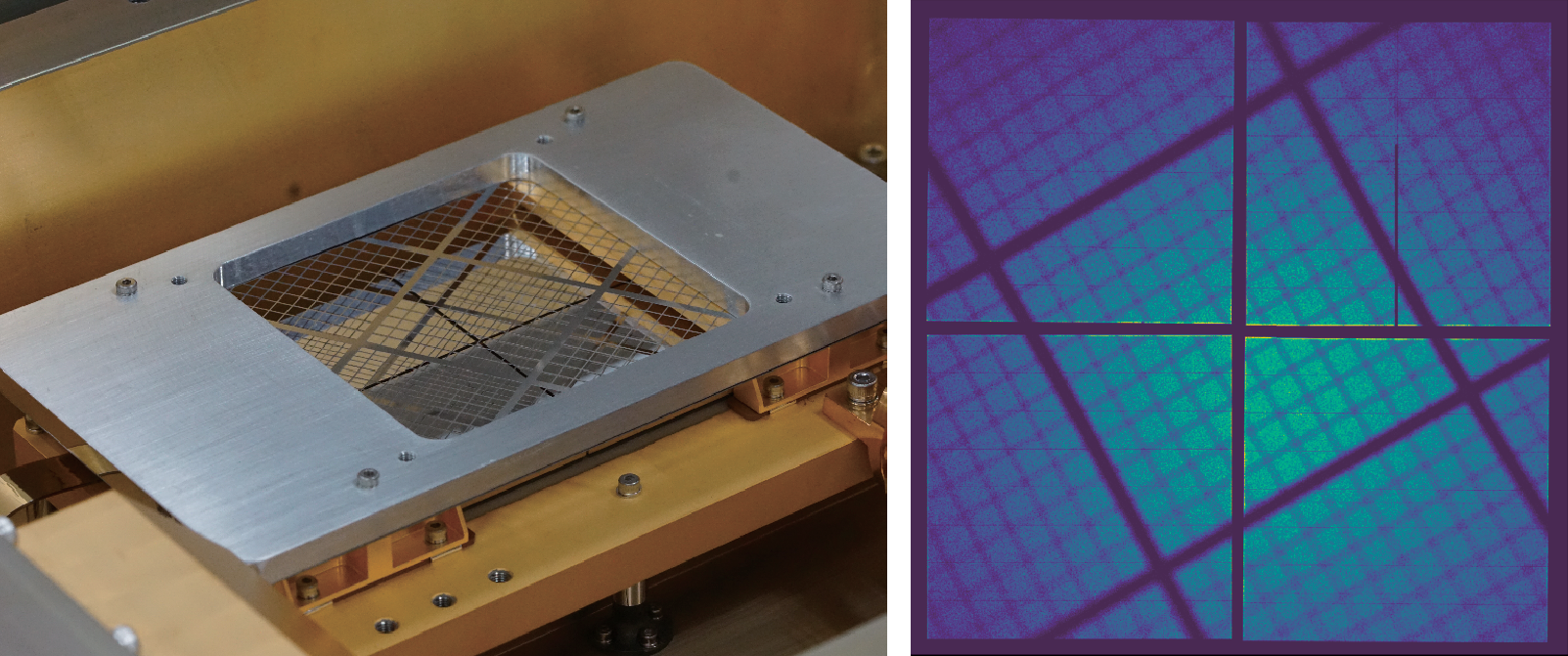}
 \end{center}
 \caption{(Left)  Photograph of the mesh mounted on the CCD chips during the SXI stand-alone cooling test. (Right) X-ray image  where  the entire imaging area is irradiated with Mn-K$\alpha$ photons through the mesh and  the orientation  between the four CCD chips are adjusted. The distinctive black lines running horizontally and vertically through the center are the CCD chip gaps. The  black lines diagonal to the horizontal axis  are  shadows of the mesh,  whereas the thin horizontal and vertical lines  are bad columns and those with artificially injected charges with  CI, respectively. {Alt text: One photograph and one X-ray image.}}
\label{fig11}
\end{figure*}
%\end{verbatim}

%------------------------------------------------------------
\subsubsection{Event extraction in SXI-DE}
%------------------------------------------------------------

The processing in SXI-DE involves several steps. 
 First, the average pulse height of the horizontally overclocked pixel values is subtracted from the pulse heights of imaging-area pixels  in every row to exclude short-term variability, which  has not been considered in the dark-level calculation at the previous step in SXI-PE. 
Next, to refine the data further,  three filters are applied: the area-discrimination, surround, and $3\times3$ local-maximum filters.
The area-discrimination filtering limits the regions in the imaging area and excludes specified regions from X-ray event searches. 
The surround filtering removes events with charges distributed  over neighboring pixels to eliminate charged-particle background events. 
The $3\times3$ local-maximum filtering identifies events with  the central pixel in a $3\times 3$ pixel region having a higher pulse height than its eight surrounding pixels. 
The filtered event data are then transmitted to the DR.  In the eventual 
%(or in some cases immediate) 
downlink  to ground,  the event location, time, and the pulse height of the $5 \times 5$ pixels of each event are sent. 
The data of the inner $3\times3$ pixel region and outer 16 pixel region of the $5\times 5$ pixel region are separated into distinct packets and may or 
 may not be sent to the ground, depending on the available downlink bandwidth, as they are flagged as lower-priority data. 
SXI-DE compresses both $3\times 3$ and $5\times 5$ event data to reduce the size of telemetry. 

%=============================
\subsection{Post processing on the ground}
%=============================

%-------------------------------------------------------------------------
\subsubsection{Even/Odd correction}
%-------------------------------------------------------------------------

As  described in \S3.1.3, the even and odd columns  are independently AD-converted in the signal chain  in MND02 on the video board to shorten the processing time. 
The ASICs for the AD conversions have slightly different gain and offset levels, which lead to a raw image  with slightly different pulse-height values and the pedestal (offset) levels  between even and odd columns. 
In addition, \citet{Nakajima2020} found that  the gain of the ASICs for the AD conversion is  affected by the temperature of the MND02 chip. 
To address this, we performed a temperature-dependent calibration of the gain and offset levels of the ASICs for the AD conversion  after  installing the ASICs onto the video boards before the launch of \textit{XRISM}. 
The calibration result indicates a temperature dependency of approximately $0.1$\%/20~$^{\circ}$C for the flight-model candidate chips,  which is incorporated into the calibration database. 
The temperature-dependent pulse-height difference between the even and odd columns  is corrected (called ``Even/Odd correction'') during the on-ground pipeline process.

%---------------------------------------------------
\subsubsection{Charge trail and CTI correction}
%---------------------------------------------------

This process addresses the charge loss during a charge transfer. 
Depending on the trap level depth, timescales of trapping and releasing signal charges  vary. 
In the case of PchNeXT4A, at least three trap populations exist with different timescales  (\cite{Kanemaru2020}). 
The shortest timescale is approximately equal to the transfer time of one logical pixel, leading to trailing charges at the pixel next to the event center pixel. 
The correction for this effect is called the charge-trail correction. 

Following the charge-trail correction, the pulse height of each event is corrected according to   the CTI with a timescale longer than the transfer period (\cite{Kanemaru2020}).
This is called the CTI correction. 
 Without the CTI correction, the raw pulse height generally decreases with the number of charge transfers, 
defined  with the formula $\textrm{PH} = \textrm{PH}_0 \times (1 - \textrm{CTI})^n$, where PH$_0$ and PH  are the pulse height amplitudes before and after charge transfers through $n$ pixels, respectively. 
 Since we employ the CI technique (see \S3.1.2) during operation, the raw pulse height shows a periodic recovery (i.e., a\ saw-tooth pattern) at the CI rows placed at every 80-th logical row (see \S5.2). 
After the CTI correction, the saw-tooth pattern disappears, and the pulse height becomes independent of the row number.  The resultant pulse-height profile is consistent with the expectation in the ideal case without a charge loss during transfer. 
It is evident that the CTI correction enhances the peak,  narrowing the width,  in the spectrum of monochromatic X-rays, improving the energy resolution.
  Since the CTI increases over time due to radiation damage in orbit, the CTI should be re-calibrated during the mission. For this purpose, the in-orbit calibration is carried out  with onboard calibration-source data and celestial target spectra. 

%---------------------------------------------------
\subsubsection{Pulse-height correction based on ``grades''}
%---------------------------------------------------

To distinguish between X-ray and particle events, we employ  a ``grade''  method, which classifies grades of events according to the distribution of the pulse height after the charge-trail and  CTI corrections. 
Specifically, %we adopt for the \textit{XRISM}/SXI the identical grade selection method  with that applied  to the \textit{Suzaku}/XIS. 
For the \textit{XRISM}/SXI, we adopt the same grade selection method as that applied to the \textit{Suzaku}/XIS.
The method uses two thresholds: the event threshold and split threshold.
If the leakage of the charge cloud to the neighboring pixels  of the event-center pixel is smaller than the split threshold, the resultant  pulse height is underestimated. 
 This effect is corrected  according to the amount of mean expected charge loss for each grade \citep{Aoki2023}.

%---------------------------------------------------
\subsubsection{Gain and line-profile correction}
%---------------------------------------------------

In this step, the corrected pulse height is converted to energy using the response function. 
The response function is mainly composed of  a gain correction and corrections based on the line profile and quantum efficiency.  
The gain correction converts the corrected pulse height into a final pulse invariant (value calibrated to align the correspondence between pulse height and energy across all CCDs),  
 using the pre-flight calibration data (see \S5.2)   in such a way that the conversion satisfies the  requirement that the difference between the energy calculated from the pulse invariant and the intrinsic energy  fall within $5\%$ at 1~keV and $0.3\%$ at 6~keV. Here, the relation between the pulse invariant value and energy is defined to be $1~\textrm{ch} = 6~\textrm{eV}$ for the \textit{XRISM}/SXI. 

The pulse-height distribution of the CCD events  of monochromatic X-rays consists of the following five components.
\begin{itemize}
\item Primary Gaussian originating in  the events  where  all the charges generated by  incident X-rays are collected. 
\item Secondary Gaussian originating in  the events  where  the generated charge cloud spreads over multiple pixels and  part of the charges are not collected. 
\item Constant component originating in the  events  where  part of the generated charge cloud spreads outside the depletion layer. 
\item Si line produced by a Si-K$\alpha$ line or an Auger electron from other pixels.  
\item Si escape, which is the remainder of the original energy  when  a Si-K$\alpha$ is emitted and its energy (1.74~keV) is lost accordingly.  
\end{itemize}
See, for example, \citet{Inoue2016} and \citet{Tanaka2018}. 

We investigated the gain and line profile  before the launch by using the data obtained with the multi-color X-ray generator \citep{Yoneyama2020} and the on-board calibration sources and constructed a response function  to account for them.   
After the launch, the response function is  updated with observation data  when necessary. 

%============S5============
\section{Pre-Flight Performance}
%============S5============

With  extensive pre-flight performance-verification experiments with the full flight configuration of the \textit{XRISM}/SXI or equivalent setups, we confirmed that the \textit{XRISM}/SXI  performed at the expected level for the imaging capability (\S5.1), spectroscopic performance (\S5.2), optical blocking performance (\S5.3), cooling and temperature control capabilities (\S5.4), and transmissivity and quantum efficiencies (\S5.5).  

%==========================
\subsection{Imaging capability}
%==========================

%\begin{verbatim}
\begin{figure}
 \begin{center}
  \includegraphics[width=80mm]{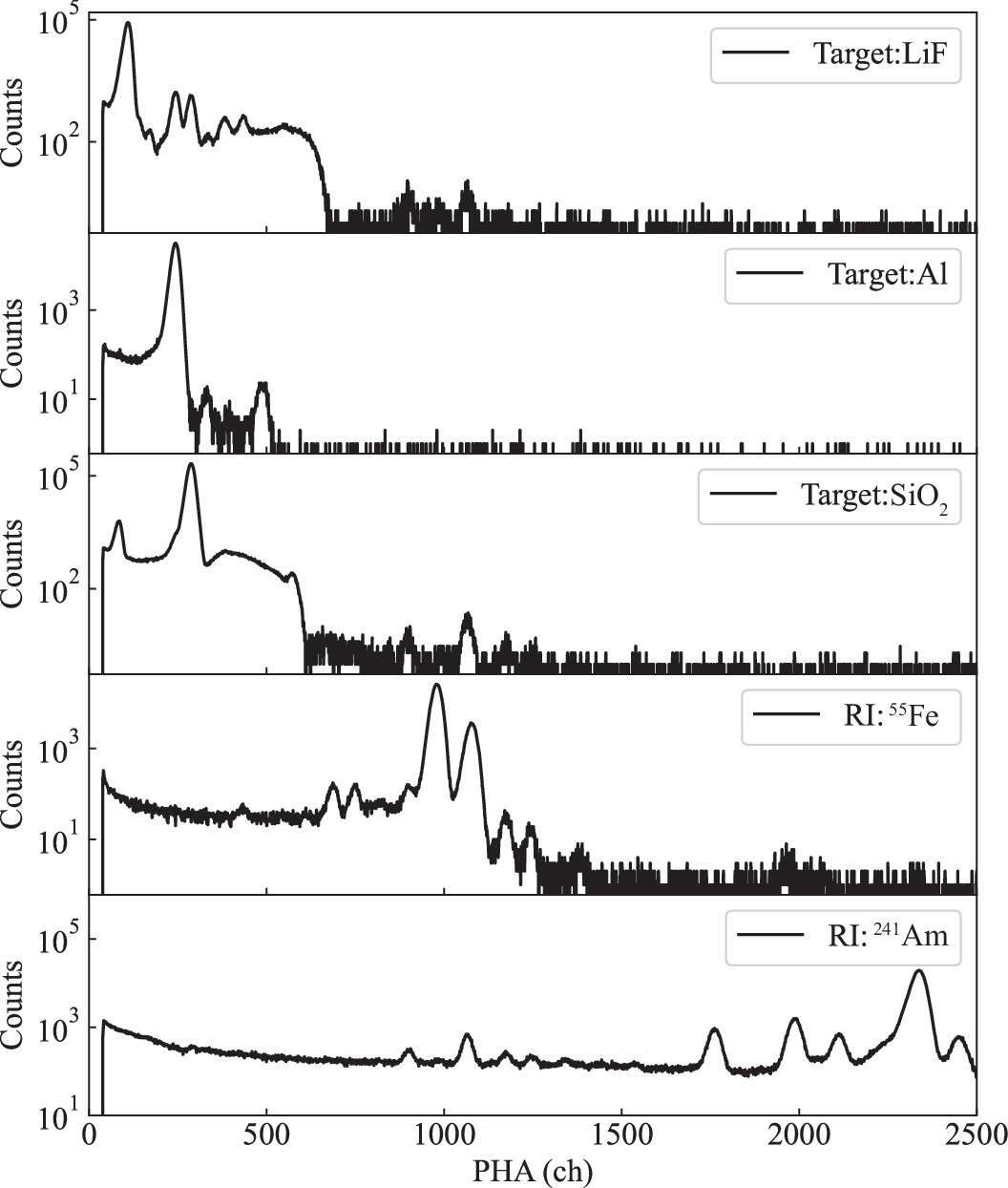}
 \end{center}
 \caption{Spectra obtained in the pre-flight calibration  with a multi-color X-ray generator. The horizontal axis is  in PHA (ch), where 1000~ch corresponds to $\sim 6$~keV. {Alt text: Five line graphs.}}
 \label{fig7}
\end{figure}
%\end{verbatim}

%\begin{verbatim}
\begin{figure*}
 \begin{center}
  \includegraphics[width=170mm]{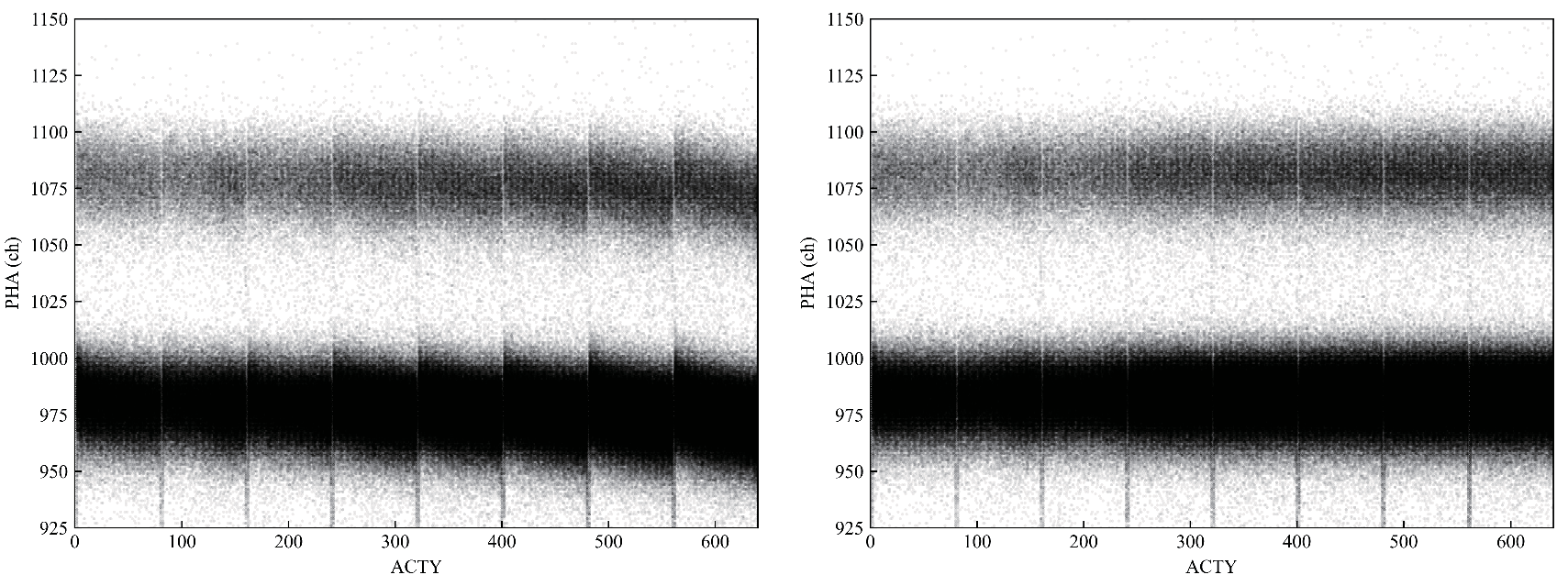}
 \end{center}
 \caption{Pulse heights of Mn-K$\alpha$ and K$\beta$ as a function of row number in the horizontal direction in a segment of a CCD chip. Left and right panels show the pulse heights before and after the charge trail and CTI corrections (\S4.2.2) were applied, respectively. {Alt text: Two scatter plots.}}\label{fig9}
\end{figure*}
%\end{verbatim}

%\begin{verbatim}
\begin{figure}
 \begin{center}
  \includegraphics[width=80mm]{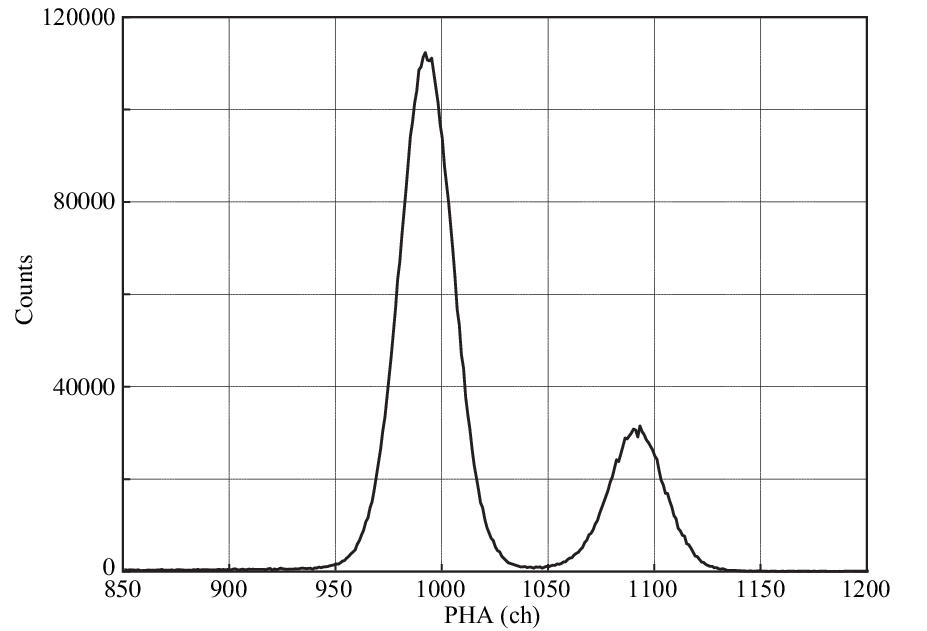}
 \end{center}
 \caption{A spectrum with a Mn-K$\alpha$ and K$\beta$  lines obtained by a segment of a CCD chip of the SXI in the full flight-model configuration. The CCD temperature was kept at $-110$~$^{\circ}$C during the data acuisition. The charge-trail and CTI corrections (\S4.2.2) have been applied. {Alt text: One line graph.}}\label{fig10}
\end{figure}
%\end{verbatim}

%\begin{verbatim}
\begin{figure*}
 \begin{center}
  \includegraphics[width=120mm]{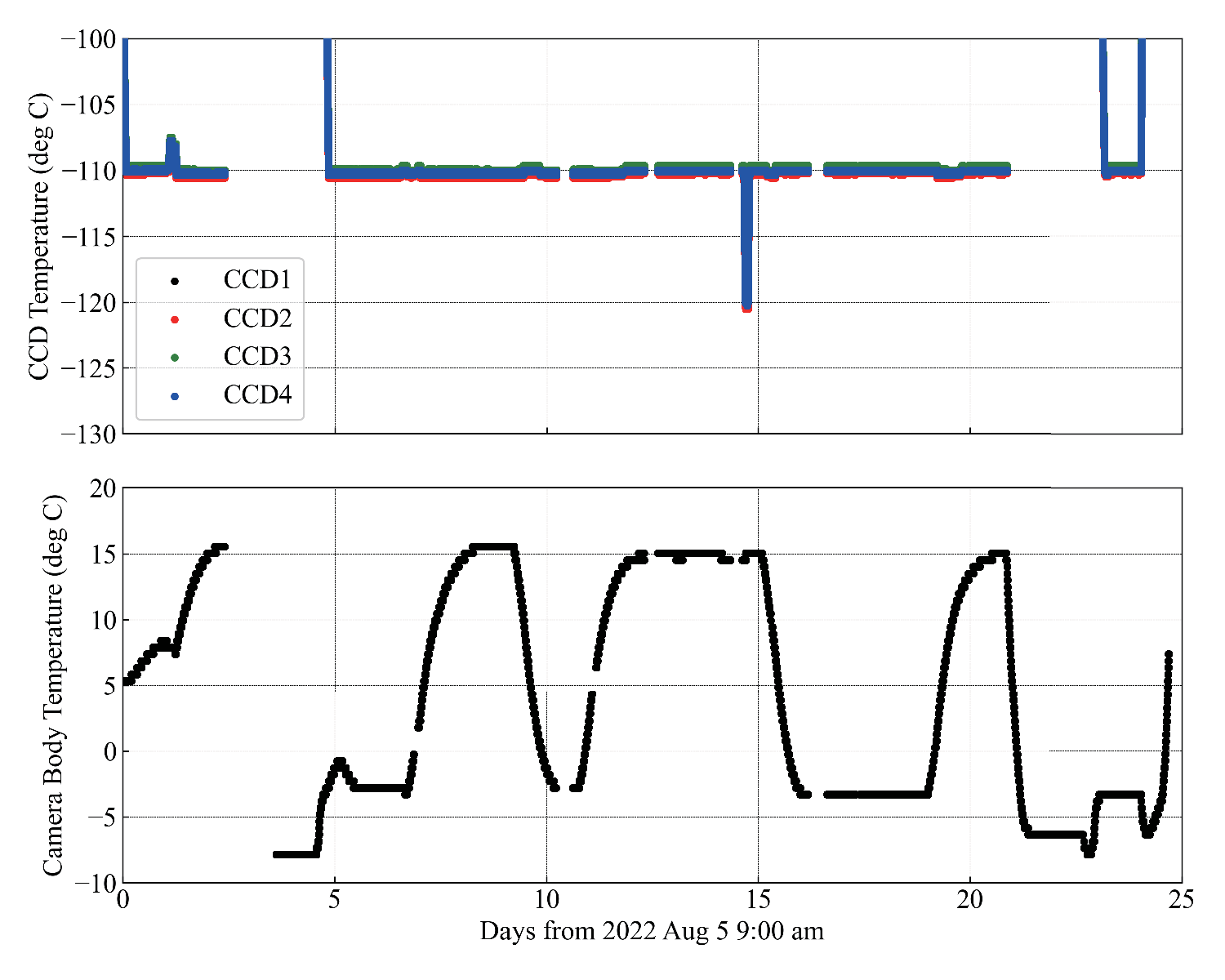}
 \end{center}
 \caption{The long-term  history of the temperatures of the (top panel) four CCD chips  and (bottom panel) camera body,  as a function of days from 9:00~am on 2022 August 5. The camera body temperature reflects the temperature of the surrounding environment  of SXI-S. The CCD temperatures first reached $-110~^{\circ}$C at $0.07$~days.  Then, the SXI body was turned at  around $1.08-1.26$~days. The CCDs were warmed up between $2.36$ and $4.55$~days, and  then cooled to $-110~^{\circ}$C at $4.85$~days. After that, the CCD temperatures  were stably kept at $-110~^{\circ}$C for $\sim 16$~days, except for  a period from 14.70 to 14.75~days,  during which the CCD temperatures were kept at $-120~^{\circ}$C. The CCDs were warmed up between $20.87$ and $22.84$~days and then cooled to $-110~^{\circ}$C at $23.18$~days. {Alt text: Two dotted graphs. }
}\label{fig14}
\end{figure*}

To  examine the imaging capability of the \textit{XRISM}/SXI, we temporarily installed a mesh frame (Figure~\ref{fig11} left) above the flight-model CCD array, and cooled it to $-110~^{\circ}$C. 
We  irradiated the entire imaging area  with 5.9-keV X-rays from a $^{55}$Fe radioisotope source  placed above the mesh.  In the experiment, 
%all  the components of the SXI-S except for SXI-S-HOOD and SXI-S-BNT were  the flight-model ones. 
the flight-model components of SXI-S were used except for SXI-S-HOOD and SXI-S-BNT.
 Figure~\ref{fig11} (right) shows the obtained X-ray image. 
We confirmed that shadows of the mesh were cast  over the expected areas  while unhindered X-rays were detected in the other areas. 
No regions on the image were insensitive to X-rays, except for bad columns and those with artificially injected charges with CI. 
Thus, we confirmed that the \textit{XRISM}/SXI had the expected imaging capability.

The CCD chips are arranged in a $2 \times 2$ mosaic configuration to achieve a wider FOV than that with a single chip. 
%Even when the four chips are arranged as densely as possible, some dead spaces between the chips cannot be avoided. 
The insensitive dead space between devices inevitably exists, and the gap must be precisely characterized.
Therefore, we needed to measure the size of these dead spaces. 
With the imaging experiment with a mesh (Figure~\ref{fig11} (right)), we determined the relative orientation of the four CCD chips. 
As a result, the gaps between the active pixel regions were found to be $1.2-1.6$~mm ($45-60''$).
 % which even our best effort of arrangement had not entirely eliminated.
\newline \indent After \textit{XRISM}/SXI and the XMA were mounted on the spacecraft, we checked the alignment between the XMA and SXI and confirmed that the on-axis position of the mirror did not fall into the gaps between the CCD chips on ground before the launch.

%To ensure that the on-axis position does not fall into the gap between the CCD chips, the alignment of the XMA and SXI should be accurately measured. 
%After \textit{XRISM}/SXI was mounted on the spacecraft, the XMA was installed. 
%Since the SXI was mounted with the opaque SXI-S-CBF, it was impossible to directly adjust the direction of the optical axis  of XMA while looking at the CCD chips. 
%Therefore, an alignment mark was placed on the cover of the SXI-S-CBF, and the optical axis of XMA was aligned using the positional relationship between this mark and the CCD chips, which had been measured in advance. 
%As a result, the focus position was offset by $(X, Y)$ = ($-8.7$~mm, $+9.3$~mm) from the center of the $2 \times 2$ CCD array, satisfying the requirements for the aim point. (この文の数字の出し方が要確認。)
%Thanks to this successful alignment of the XMA and SXI, Xtend achieved a wide FOV of $38' \times 38'$, ensuring that the on-axis position never falls into the gaps between the CCD chips.

%==========================
\subsection{Spectroscopic performance}

To measure the spectroscopic performance of the flight-model CCD chips at multiple energies, we conducted experiments using  a multi-color X-ray generator \citep{Yoneyama2020}. 
The X-ray generator includes secondary targets of LiF (F-K$\alpha$ at 0.677~keV), Al (Al-K$\alpha$ at 1.49~keV), SiO2 (O-K$\alpha$ at 0.525~keV and Si-K$\alpha$ at 1.74~keV), as well as radioisotopes such as $^{55}$Fe (Mn-K$\alpha$ at 5.9~keV) and $^{241}$Am (Np-L$\alpha$ at 14.1~keV). In the experiments, we cooled and operated the CCD chips with a non-flight-model read-out system. 
Figure~\ref{fig7} displays  a few example X-ray spectra obtained from five species of  targets  or radioisotopes.
 As a representative case, the fourth panel in Figure~\ref{fig7} shows  that the Mn-K$\alpha$ and Mn-K$\beta$ lines 
 from the $^{55}$Fe radioisotope were clearly resolved.
 %were  clearly resolved when the $^{55}$Fe radioisotope, which emits the Mn-K lines, was used for  X-ray irradiation. 
As such, we  confirmed that the SXI resolved the emission lines in the range of $\sim 0.5-14.1$~keV as expected without a significant loss of the soft X-ray sensitivity. 

Next, we changed the configuration to the full fight-model one,  and  examined the CTI and energy resolution with the $^{55}$Fe radioisotope. 
Figure~\ref{fig9} (left) shows the pulse-height values of Mn-K$\alpha$ and K$\beta$ as a function of row number in the horizontal direction in a segment of a CCD chip,   where 
 the even/odd correction (\S4.2.1) was made but not  the charge-trail  or CTI corrections (\S4.2.2).  
 The saw-tooth pattern and the decreasing trend   toward higher row numbers in the obtained distribution  are due to the lack of the latter two corrections.  
Figure~\ref{fig9}\ (right), in contrast, shows the pulse heights where all three corrections  were applied. 
The saw-tooth pattern caused by the  lines with CI in every 80 rows (Figure~\ref{fig:frame}) were corrected to some  degree, and the pulse-height values of Mn-K$\alpha$ and K$\beta$  were found to be distributed around constant values, i.e., the systematic trends in Figure~\ref{fig9} (left)  were largely eliminated.

 After  mounting the \textit{XRISM}/SXI  to the spacecraft,  we tested  its spectroscopic performance. Figure~\ref{fig10} shows the Mn-K$\alpha$ and K$\beta$ spectra obtained by a segment of a CCD chip with the $^{55}$Fe calibration source. 
We  measured  the energy resolution at 5.9~keV  to be $173-188$~eV (FWHM), which meets the requirement of $< 200$~eV (FWHM) at the BOL (see \S2). 
 With  separate experiments to examine the radiation-damage  effects, it has also been demonstrated that the CCD chips should meet the requirement of $< 250$~eV (FWHM) even after 3 years of operation in orbit (until the EOL),  mainly owing to the new notch structure introduced in the CCD chips to reduce the degree of degradation \citep{Kanemaru2019}.

%%\begin{verbatim}
%\begin{figure}
% \begin{center}
%  \includegraphics[width=70mm]{figure/response.eps}
% \end{center}
% \caption{A schematic line profile obtained by an X-ray CCD when a monochromatic X-ray is observed. }\label{response}
%\end{figure}
%%\end{verbatim}
%

%==========================
\subsection{Optical blocking performance}
%==========================

%In the flight-model CCD cooling experiments, we set a light emitting diode (LED) near the CCD chips to test the optical light leakage. 
%As a result, even the most-edge effective pixels which are most highly affected by light leakage meet the CCD light reduction of $< 10^{-4}$. 
%Furthremore, the number of pinhole pixels was confirmed to be less than 1\% on all surfaces and no significant aging degradation has occurred as far as can be confirmed \citep{Uchida2020}.
%
%The optical blocking perfromacne was evaluated by irradiating optical photons from LEDs through SXI-S-CBF in the full flight-model configuration. 
%As a result, the light leakage on the edges and pinholes was confirmed to be $<10^{-7}$. 
%Therefore, the total light leakage through SXI-S-CBF and the CCD optical blocking layer was estimated to be $<10^{-11}$ which meets the requirement of the optical blocking performance. 

During the  cooling experiments with the flight-model CCDs, we placed a light-emitting diode (LED) near the CCD chips to test  the amount of optical light leakage. 
 We found that even the outermost effective pixels, which are most affected by light leakage from the edges, met the  light-reduction requirement for the CCD of less than $10^{-4}$. 
We confirmed  that the number of pinhole pixels was  less than 1\% across  the entire surface as we had designed. No significant  degradation by aging was observed by \citet{Uchida2020}.

After the installation of SXI-S-CBF  above  SXI-S for the full flight-model configuration, we evaluated the optical blocking performance  by irradiating optical photons from LEDs through the SXI-S-CBF. 
The light leakage was  found to be less than $10^{-3}$. 
 Consequently, the total light leakage through the SXI-S-CBF and the CCD optical blocking layer was estimated to be less than $10^{-7}$, which meets the optical blocking performance requirement.

%==========================
\subsection{Cooling and temperature control capability}
%==========================

%After installing the full-flight configuration of SXI onto the spacecraft, we performed the continuous operation of SXI-S-1ST and tested the cooling and temerature controlling capability for about one month in the satellite thermal vacuum test.
%In this experiment, the spacecraft was put into a thermal vacuum chamber and the environmental temperature was varied to simulate the cold and hot cases in the satellite orbit. 
%The stability of the cooling and themperature control of the CCD chips was tested in this experiment. 
%Figure~\ref{fig14}(top) shows the long-term trends of the tempertures of the four CCD chips, and the cold head temperature that reflects the environmental temperature. 
%Except for cooling from and warming to room temperture, the CCD tempertures were confirmed to be stably kept at $-110^{\circ}$C or $-120^{\circ}$C even though the emvironmental temperature varied by $\sim 25^{\circ}$C. 
%As a result, we successfully confirmed sufficient cooling and temperature control capability with the single Striling cooler system. 

After  mounting the full-flight configuration of the SXI onto the spacecraft, we continuously operated the SXI-S-1ST and  examined the cooling and temperature control capabilities for about one month during the satellite's thermal vacuum test. 
In this experiment, the spacecraft was placed in a thermal vacuum chamber, and the environmental temperature was varied to simulate both the cold and hot conditions that the satellite would experience in orbit.
The stability of the cooling system and the temperature control of the CCD chips were evaluated in this experiment. Figure~\ref{fig14} shows the long-term temperature  history of the four CCD chips during the experiment, along with the camera body temperature\footnote{The temperature was measured at the point  between the compressor of SXI-S-1ST and the cooler dummy.}, which reflects the environmental temperature. 
Except for the cooling and warming phases, 
the CCD temperatures were confirmed to remain stable at the set temperatures of either $-110~^{\circ}$C or $-120~^{\circ}$C, even  though the environmental temperature change was  approximately $22~^{\circ}$C.
 Thus, we successfully verified the sufficient capability of the SXI for cooling and temperature control  with the single Stirling cooler system (\S3.1.5).

%==========================
\subsection{Detection efficiency}
%==========================

%In \textit{XRISM}/SXI, the transmissivity of SXI-S-CBF and the quantum efficiency of the CCD chips mainly affect the effective area of Xtend, because X-ray photons from objects directly transparents SXI-S-CBF and enter the CCD chips without going through other components. 
%To measure them, we performed an X-ray beam experiment at Photon Factory of High Energy Accelerator Research Organizaion (KEK). 
%The CBFs used in the experiment are the same as the flight model except that it does not have a mesh structure with an aperture ratio of 0.892 to support the structure.
%The CCD chips used in the experiment are mini-size CCD elements with the equivalent structures to those of flight models. 
%Figure~\ref{fig12}(left) shows the measured transmissivity of SXI-S-CBF, depending on incident energies, and the model function that reproduces the measurements. 
%The transmissivity was meausred to be 0.857 at 1.5~keV and 0.892 at 6~keV after taking the presence of the mesh structure into account. 
%Figure~\ref{fig12}(right) shows the measured quantum efficiency of the CCD chips and the model to reproduce the experiment. 
%The quantum efficiency of the CCD chips was measured to be 0.957 at 1.5~keV and 0.993 at 6~keV. 
%As a result, we successfully confirmed that they meet the allocations of the transmissivity and the quantum efficiency to satisfy the effectve area of Xtend ($270$~cm$^2$ at 6~keV and $300$~cm$^2$ at 1.5~keV). 

The detection efficiency of the \textit{XRISM}/SXI is defined  as a multiplication of  the transmissivity of  SXI-S-CBF and the quantum efficiency of the CCD chips. 
The effective area of Xtend is basically defined as the multiplication of the detection efficiency of the SXI and the effective area of the XMA because X-rays interact with only the XMA and SXI-S-CBF before entering the CCD chips.
%because the physical process for X-ray photons from observed objects to be detected is simple; the photons are first reflected by the XMA, pass directly through  SXI-S-CBF, and then enter the CCD chips without interacting  any other components. 
To evaluate the detection efficiency, we conducted an X-ray beam experiment at the Photon Factory of the High Energy Accelerator Research Organization (KEK), where we separately measured
the transmissivity of  SXI-S-CBF and the quantum efficiency of the CCD chips. 
The CBFs used in this experiment were identical to the flight models, except for the absence of a mesh structure with an aperture ratio of 0.896, which supports the structure. 
The CCD chips used in  this experiment were  smaller  models than the flight models but with structures equivalent to those of the flight models.

\begin{figure*}
 \begin{center}
  \includegraphics[width=160mm]{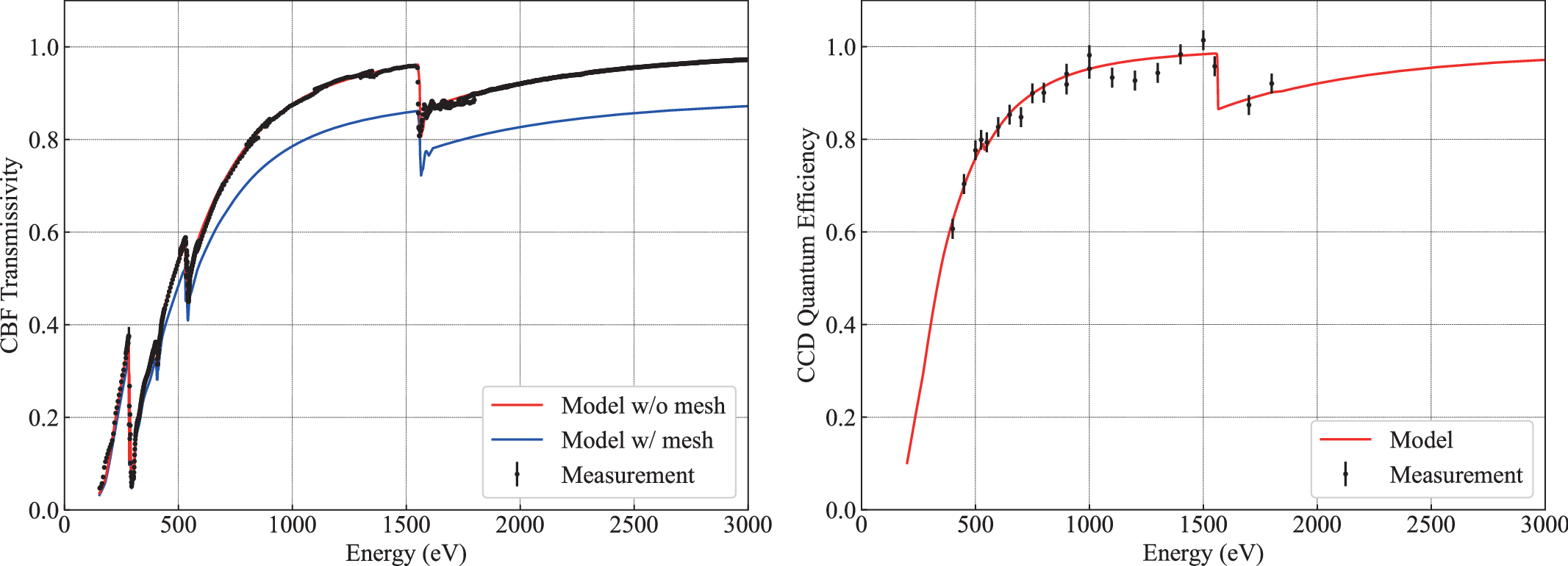}
 \end{center}
 \caption{ Model and measured values for the (Left panel) Transmissivity of SXI-S-CBF  and  (Right panel) quantum efficiency of the CCD chips, as a function of incident energy. {Alt text: Two line graphs.}}\label{fig12}
\end{figure*}

Figure~\ref{fig12}~(left) shows the measured transmissivity of the SXI-S-CBF as a function of incident energy, along with a (best-fit) model that reproduces the measurements. We also overlay in the figure the model transmissivity curve for the real flight model with the mesh, calculated from the best-fit model and physical parameters of the mesh. 
The transmissivity for the real flight model was  estimated to be 0.857 at 1.5~keV and 0.892 at 6keV. 
From the measured transmissivity, we calculated the thicknesses of the Al and polyamide layers   to be $102.3 \pm 0.5$ and $238.2 \pm 0.5$~nm, respectively,  and that of an Al$_2$O$_3$ layer  to be $7.5\pm 0.5$~nm. 
Figure~\ref{fig12}~(right) presents the measured quantum efficiency of the CCD chips, along with a model that  reproduces the experimental results. 
The quantum efficiency of the CCD chips was measured  to be 0.957 at 1.5~keV and 0.993 at 6~keV.
From the quantum efficiency, the thickness of the Al layer of the CCD chip was  estimated to be $230 \pm 8$~nm  under assumptions of the thicknesses of the SiO$_2$ and Al$_2$O$_3$ layers  being 20 and 0~nm, respectively. 
As a result, we  confirmed that both the transmissivity and quantum efficiency meet the  requirements to achieve the effective area of Xtend in its specifications, $270$~cm$^2$ at 6~keV and $300$~cm$^2$ at 1.5~keV.

%\begin{verbatim}
%\end{verbatim}

%==========================
\section{Conclusion}
%==========================

Xtend is the soft X-ray imaging telescope onboard \textit{XRISM}, which was adopted as a recovery mission for \textit{Hitomi} and whose design is broadly based on that of \textit{Hitomi}. 
We refined the requirements for Xtend, and accordingly designed and developed the X-ray CCD camera for Xtend, the Soft X-ray Imager (SXI), which covers  an energy range of 0.4--13.0~keV. This paper presents its design and performance which we verified with pre-flight tests. 
For the SXI, we employed back-illuminated CCDs with a 200-$\mu$m-thick depletion layer, the same as those onboard \textit{Hitomi}, but with some significant  improvements in the optical blocking capability and charge transfer efficiency.  
  The SXI has the $2 \times 2$ CCD array and achieves, in combination with the XMA, a large field of view (FOV) of $38' \times 38'$ (large enough to cover the full moon), which encompasses the FOV of the soft X-ray spectroscopy telescope, Resolve, the other instrument of \textit{XRISM}.
We conducted pre-flight experiments with the full flight-model configuration of the SXI and  confirmed that the imaging capability and spectroscopic performance of the \textit{XRISM}/SXI met expectations and  satisfied the requirements for Xtend. 
We also  examined  the optical blocking performance, cooling and temperature control capabilities, and detection efficiency of the SXI,  using  its full flight configuration or equivalent setups, and found that all of them met the requirements.
Given these results, the SXI of \textit{XRISM} can be considered one of the best X-ray CCD cameras onboard X-ray astronomical satellites, 
%following the X-ray Imaging Spectrometer (XIS) \citep{Koyama2007} onboard the Japanese fifth X-ray astronomical satellite \textit{Suzaku}  
in the sense that it has  a relatively large FOV with a stable NXB level  for  a focal-plane X-ray detector, and the combination with the XMA provides the largest grasp at 7~keV among the X-ray CCD cameras onboard large observatories. 
\textit{XRISM} was successfully launched in 2023, and since then, the SXI, too, has been operated in orbit. The in-orbit performance of Xtend, including whether the effective area and NXB levels meet the requirements, will be reported in a subsequent paper.

\begin{ack}
The authors thank Hiromichi Okon, Takuto Narita, Kai Matsunaga, Yujiro Saito, Moe Anazawa, Hirotake Tsukamoto, Honoka Kiyama, Kaito Fukuda, Mariko Saito, Shuusuke Fudemoto, Satomi Onishi, Junichi Iwagaki, Kazunori Asakura, Maho Hanaoka, Yuichi Ode, Tomohiro Hakamata, Mio Aoyagi, Shunta Nakatake, Toshiki Doi, Kaito Fujisawa, and Mitsuki Hayashida for their contribution to the development of the \textit{XRISM}/Xtend/SXI.
This work is supported by Japan Society for the Promotion of Science (JSPS) KAKENHI with the Grant number of 19K21884, 20H01947, 20H01941, 23K20239 (H.N.), 23K20850, 21H01095 (K.M.), 20KK0071, 24H00253 (H.N.), 19K03915, 23K22536, 24K21547 (H.U.), 21J00031, 22KJ3059, 24K17093 (H.S.),  21K03615, 24K00677 (M.N.), 20H00175, 23H00128 (H.M.), 21H04493, 15H02090, 14079204 (T.G.T.), 22H01269 (T.K.), 24K17105 (Y.K.). 
K.~K.~N. acknowledges the support by the Yamada Science Foundation. 
\end{ack}

\end{document}